\documentclass[conference]{IEEEtran}
\IEEEoverridecommandlockouts
\usepackage{cite}
\usepackage{amsmath,amssymb,amsfonts}
\usepackage{algorithm}
\usepackage{algorithmic}

\usepackage{graphicx}
\usepackage{textcomp}
\usepackage[table]{xcolor}
\usepackage{multirow}
\usepackage{makecell}
\usepackage{booktabs}
\usepackage{array}
\usepackage{subcaption}
\usepackage{url}
\usepackage[colorlinks=true,citecolor=blue,linkcolor=blue,urlcolor=blue]{hyperref}
\usepackage{stfloats}
\usepackage[misc]{ifsym} 

\def\BibTeX{{\rm B\kern-.05em{\sc i\kern-.025em b}\kern-.08em
    T\kern-.1667em\lower.7ex\hbox{E}\kern-.125emX}}

\DeclareRobustCommand*{\IEEEauthorrefmark}[1]{%
    \raisebox{0pt}[0pt][0pt]{\textsuperscript{\footnotesize\ensuremath{#1}}}}

\begin{document}

\title{RISE: Relay Inference and Online Scheduling \\ for Efficient Edge-Device Collaborative \\Diffusion Model Services}
\author{
\thanks{This work was supported in part by the National Natural Science Foundation of China (NSFC) under Grant 62302048, Grant 62272050, and Grant U25A20436; in part by the Science and Technology Development Fund of Macau SAR under Grants 0028/2025/AFJ and 0021/2025/RIA1; in part by Guangdong Higher Education Association under Grant 24GQN97; in part by the Guangdong Provincial Higher Education Institutions under Grant 2024KTSCX219; and in part by Beijing Normal University at Zhuhai Education Reform Project under Grant jx2025037.} \thanks{\textit{(Corresponding author: Zhiqing Tang.)}}
\IEEEauthorblockN{
Zilan Huang\IEEEauthorrefmark{1,2},
Zhiqing Tang\IEEEauthorrefmark{2,1}$^{\textrm{\Letter}}$,
Hanshuai Cui\IEEEauthorrefmark{3,2},
Tian Wang\IEEEauthorrefmark{2},
Yuan Wu\IEEEauthorrefmark{4},
Weijia Jia\IEEEauthorrefmark{2,5},
Wei Zhao\IEEEauthorrefmark{6}
}
\IEEEauthorblockA{
\IEEEauthorrefmark{1}{Faculty of Arts and Sciences, Beijing Normal University, Zhuhai, China}
}
\IEEEauthorblockA{
\IEEEauthorrefmark{2}{Institute of Artificial Intelligence and Future Networks}, {Beijing Normal University}, Zhuhai, China}
\IEEEauthorblockA{
\IEEEauthorrefmark{3}{School of Artificial Intelligence}, {Beijing Normal University}, Beijing, China
}
\IEEEauthorblockA{
\IEEEauthorrefmark{4}{State Key Laboratory of Internet of Things for Smart City}, {University of Macau}, Macau SAR, China
}
\IEEEauthorblockA{
\IEEEauthorrefmark{5}{Guangdong Key Lab of AI \& Multi-Modal Data Processing}, {Beijing Normal-Hong Kong Baptist University}, Zhuhai, China
}
\IEEEauthorblockA{
\IEEEauthorrefmark{6}{Faculty of Computer Science and Artificial Intelligence}, {Shenzhen University of Advanced Technology}, Shenzhen, China
}
Email: \{zilanhuang, hanshuaicui\}@mail.bnu.edu.cn,
\{zhiqingtang, tianwang, jiawj\}@bnu.edu.cn,\\yuanwu@um.edu.mo, zhaowei@suat-sz.edu.cn
}

\maketitle

\begin{abstract}
Text-to-image diffusion models are increasingly deployed at the network edge to serve heterogeneous workloads with diverse quality and latency requirements. However, existing deployment strategies choose either large edge-side models with high fidelity but high latency or lightweight device-side models that offer speed at the cost of semantic coherence. Moreover, these approaches rarely split the denoising workload between models of different sizes across edge servers and user devices. To bridge this gap, we propose RISE, a method for edge-device diffusion model services that combines relay inference with online scheduling. Driven by the finding that the latent intensity exhibits minimal deviation after a model handoff, RISE uses a training-free relay mechanism that exploits the shared latent space within a model family: the large model on the edge handles the early denoising steps that shape semantic structure, then passes the intermediate latent to a small device-side model for detail refinement. To deploy this mechanism as a practical service, a contextual bandit scheduler selects the best relay configuration based on prompt complexity, user preferences, network quality and real-time node loads. Experiments on two benchmarks show that RISE's relay mechanism achieves up to 2.1$\times$ speedup while preserving full-model quality, and its context-aware scheduler effectively balances quality and latency under mixed workloads.
\end{abstract}


\section{Introduction}

% --- P1: Broad Context ---
Collaboration between large and small models has become a key research direction in service computing \cite{li2026toward}. In natural language processing, techniques such as speculative decoding \cite{hu2025specsurvey,yan2025speculative,ning2025dssd} have shown that pairing a large model with a smaller one can reduce cost while keeping output quality high. For text-to-image generation, diffusion models now follow a similar pattern. Model families like SDXL \cite{podell2023sdxl} and Stable Diffusion 3 \cite{esser2024sd3} offer both a large server-grade variant and a compact version designed for lighter hardware, and models within the same family share a compatible latent space \cite{rombach2022ldm,podell2023sdxl}. This compatibility gives a natural basis for collaboration between the two scales. Most prior studies focus on speeding up a single model \cite{geng2025ect,chen2025snapgen} or distributing one model across identical devices \cite{jhoo2025pfeife}. However, few studies explore the collaboration between a large and a small diffusion model deployed on an edge server and a user device, where the two sides differ significantly in computing power yet both generation quality and inference speed must be maintained.

% --- P2: Pain Points + Gap + Bridge ---
Current approaches to deploy diffusion models can be roughly divided into three types, with each of them suffering some drawbacks. \textbf{(i) Edge-only inference} places a large model such as SD3.5 Large on GPU-equipped edge servers. The output quality is high, but every image incurs high latency and heavy GPU usage \cite{zheng2025diffusion}, making it hard to serve latency-sensitive requests. \textbf{(ii) Device-only inference} runs a lightweight model such as Segmind-Vega~\cite{gupta2024progressive} directly on the user's device. Network transfer is avoided and prompt privacy is preserved, but the limited capacity of these models often produces semantic errors and visible artifacts when the input prompt is complex \cite{ben2025mobile}. \textbf{(iii) Existing collaborative inference} approaches, including split inference for CNNs and LLMs \cite{korol2025iot}, relay diffusion across different resolutions \cite{teng2024relay,zheng2024cogview3}, and cloud-edge collaboration for LLM serving \cite{jin2025icws}, are either confined to a single device or run inside a cluster of identical machines, and do not split the denoising workload between an edge server and a user device that have very different computing capabilities. Beyond single-request inference, a practical service should be able to handle many users at the same time, routing each request to a suitable node under changing loads. Relay inference and online scheduling for edge-device diffusion model services therefore remain open problems, and two challenges should be addressed.

\begin{figure*}[t]
    \centering    \includegraphics[width=0.85\textwidth]{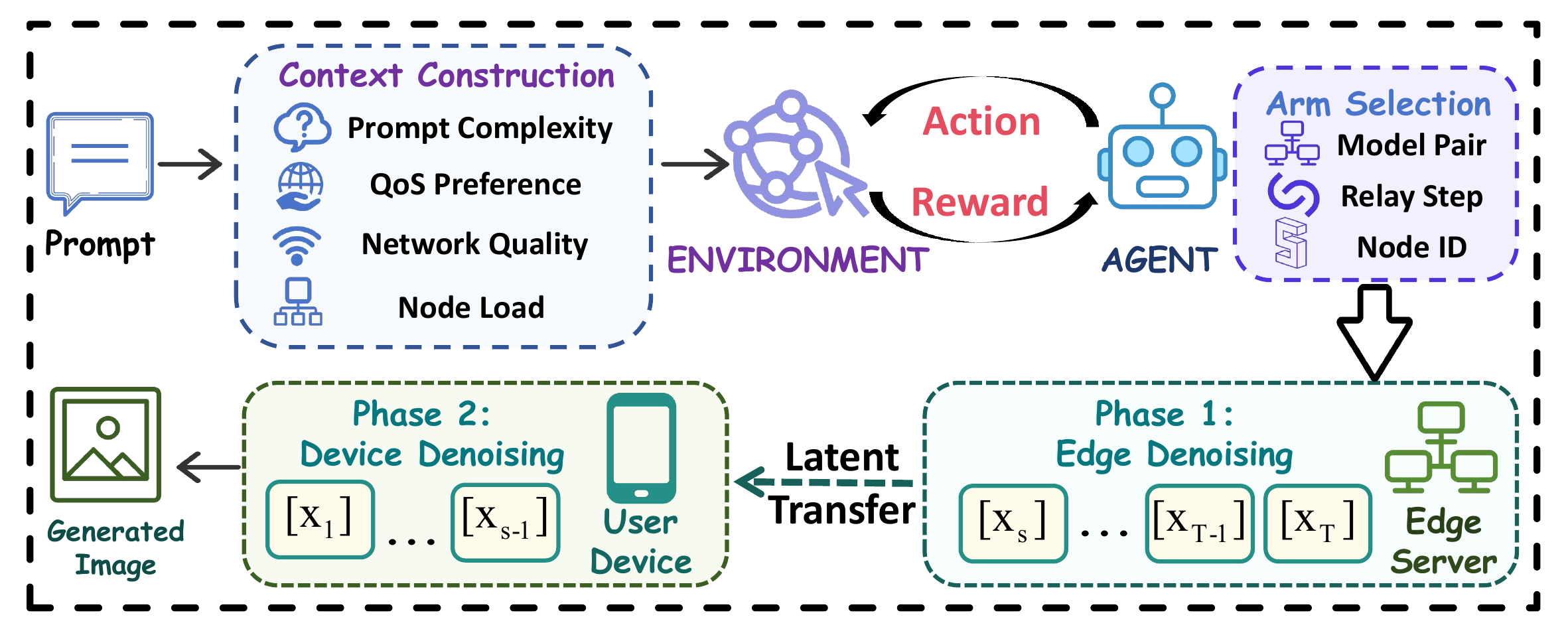}
    \caption{System overview of RISE integrating relay inference and online scheduling}
    \label{fig:system}
\end{figure*}

% --- P3: Challenge 1 (separate paragraph, italic, 5-part) ---
\textit{The first challenge is how to split a denoising task between a large edge model and a small device model without sacrificing quality.} In diffusion models, different denoising steps have different roles, and handing off at the wrong step can break the image's semantic coherence~\cite{choi2022p2,liu2024tgate}. Existing relay diffusion methods \cite{teng2024relay,zheng2024cogview3} pass a low-resolution result into a high-resolution diffusion chain, base-refiner pipelines such as SDXL run two models in sequence on the same machine, and eDiff-I \cite{balaji2022ediffi} assigns different expert networks to different noise levels. However, all of these methods either relay between fixed resolution stages or within a predefined model pair on identical hardware, and none of them hands the denoising work from a large edge model to a smaller device model of a different capacity. How to determine a valid relay point across different noise schedules and step counts while controlling quality degradation remains challenging.

% --- P4: Challenge 2 (separate paragraph, italic, 5-part) ---
\textit{The second challenge is how to select the best relay configuration for each incoming request under changing system loads.} Different model pairs combined with different relay steps create a large discrete configuration space, and the best choice depends on prompt complexity, user quality-latency preference, network quality and real-time node occupancy. Concurrent requests also compete for shared edge GPU resources, which makes the overall optimization NP-hard \cite{yang2025online}. Existing edge scheduling methods model the problem as mixed-integer programs \cite{xu2025enhancing} or apply Lyapunov optimization \cite{verma2025lyapunov}, but the former scales poorly and the latter assumes convex cost functions that do not hold for the discrete quality-speed tradeoffs in image generation. Deep reinforcement learning methods such as PPO \cite{schulmanproximal} and SAC \cite{haarnoja2018sac} have also been used for edge task offloading, but they require large amounts of training data and the learned policies are hard to interpret. Even recent QoS-aware AIGC scheduling work \cite{xu2026eat} focuses on throughput for a single model rather than jointly choosing a model pair, a relay step, and a target node. None of these methods captures the relay-specific quality-latency tradeoff unique to diffusion generation, and efficiently scheduling relay configurations in this setting remains an open problem.

% --- P5: Proposed Method (condensed, no repetition with challenges) ---
To address these two challenges, we propose RISE, a framework denoting \textbf{R}elay \textbf{I}nference and online \textbf{S}cheduling for efficient \textbf{E}dge-device collaborative diffusion model services. For the first challenge, RISE exploits the observation that early denoising steps build coarse semantic structure while later steps only add fine details \cite{choi2022p2,liu2024tgate}. Since models within the same family share an identical latent space, the large edge model can complete the early semantically important steps and pass the intermediate latent directly to the small device model, which finishes the remaining detail refinement without any retraining. For the second challenge, RISE uses a LinUCB-based contextual bandit scheduler that encodes prompt complexity, user preference, network quality and real-time node loads into a context vector and selects the best relay configuration per request.

% --- P6: Contributions ---
In summary, this paper makes the following contributions.
\begin{enumerate}
    \item \textbf{Relay Inference.} We propose a training-free relay inference mechanism for edge-device diffusion models that splits the denoising process between a large edge model and a small device model. It achieves significant speedup while preserving most of the large model's generation quality, without any retraining or fine-tuning.

    \item \textbf{Online Scheduling.} We model the joint selection of model pair, relay step, and target node as a contextual bandit problem and develop a scheduling algorithm that adapts to prompt complexity, user preferences, network quality and changing edge loads.

    \item \textbf{Experimental Validation.} We conduct experiments on two benchmarks with five complementary metrics, demonstrating that our relay mechanism achieves up to $2.1\times$ speedup while preserving full-model quality, and our online scheduler effectively balances quality and latency under mixed service workloads.
\end{enumerate}

The remainder of this paper is organized as follows. Section~\ref{sec:related} reviews related work. Section~\ref{sec:relay} presents the relay inference mechanism. Section~\ref{sec:algorithm} describes the online scheduling algorithm. Section~\ref{sec:experiment} reports the experimental results. Section~\ref{sec:conclusion} concludes the paper.

\section{Related Work}
\label{sec:related}

\subsection{Diffusion Model Acceleration}

Most research on accelerating diffusion models focuses on a single model. On the training side, trajectory stitching \cite{pan2025tstitch}, inference-time distillation \cite{park2025itdd}, and consistency tuning \cite{geng2025ect} reduce the number of denoising steps required. Training-free approaches also exist: Denoising Diffusion Implicit Models \cite{song2021ddim} replaces the stochastic sampler with a deterministic one; Latent Diffusion Models \cite{rombach2022ldm} move the denoising loop into a smaller latent space; and cache-based methods such as DeepCache and BWCache reuse intermediate features to reduce redundant computation during denoising~\cite{ma2024deepcache,cui2025bwcache}. These efforts have produced models spanning a wide range of scales, from large server-level models such as SDXL \cite{podell2023sdxl} and Stable Diffusion 3 \cite{esser2024sd3} to lightweight on-device models. They all share one limitation, however: they optimize within a single model and do not consider splitting diffusion workloads across an edge-device architecture where a large model and a small model collaborate.

\subsection{Collaborative Inference and Relay Diffusion}

Running diffusion models outside centralized data centers has motivated several collaborative inference strategies, including conditional diffusion under noisy conditions \cite{xu2025icws}, collaborative inference in vehicular networks \cite{zheng2025icws}, and cloud-edge LLM serving \cite{jin2025icws}. The iterative nature of diffusion denoising further enables cross-model splitting. eDiff-I \cite{balaji2022ediffi} assigns different expert networks to different noise levels, SDXL adopts a two-stage base-refiner pipeline, and Relay Diffusion \cite{teng2024relay} together with CogView3 \cite{zheng2024cogview3} relay across resolutions in latent space. These relay methods, however, all operate on a single machine or a cluster of identical nodes. None of them transfer the denoising workload from a large edge model to a smaller device model with different capacity, which is the setting our relay inference mechanism targets.

\subsection{Inference Scheduling on Heterogeneous Edge Clusters}

Inference scheduling has been extensively studied. Classical approaches formulate it as a mixed-integer program \cite{pournazari2025offload,xu2025enhancing} or apply Lyapunov optimization \cite{verma2025lyapunov}, while heuristic methods \cite{sultana2025hadar,wang2025optimedgeai}       trade optimality for better scalability. Deep reinforcement learning has also been applied in this area, for example, PPO \cite{schulmanproximal} and SAC \cite{haarnoja2018sac} have been used for cluster resource management, and multi-armed bandit algorithms \cite{li2018linucb,auer2002ucb} provide a more lightweight alternative with provable regret bounds. Recent work has begun exploring QoS-aware scheduling and bandit-based formulations for AIGC and LLM workloads at the edge \cite{xu2026eat,li2025bandit,yao2025enhancing}. These methods, however, all optimize for a single model and do not jointly select a model pair, a relay step, and a target node. RISE fills this gap with a LinUCB-based scheduler operating over a discrete action space that encodes all three decisions.

\section{Proposed Methodology}
\label{sec:relay}

As illustrated in Figure~\ref{fig:system}, RISE operates on a heterogeneous edge-device system where the edge hosts large models $\mathcal{M}_L$ (e.g., SDXL, SD3.5 Large) and the device runs lightweight models $\mathcal{M}_S$ (e.g., Segmind-Vega, SD3.5 Medium). This section first presents the empirical motivation for splitting the denoising workload between $\mathcal{M}_L$ and $\mathcal{M}_S$, then details the relay implementation for each model family. Table~\ref{tab:notation} summarizes the key notations used throughout this paper.

\begin{table}[t]
\renewcommand{\arraystretch}{1.2}
\caption{Summary of Key Notations}
\label{tab:notation}
\centering
\begin{tabular}{@{}c p{5.5cm}@{}}
\toprule
\textbf{Symbol} & \textbf{Description} \\
\midrule
$x_0,\, x_t,\, x_T$ & Clean image, noisy latent at step $t$, pure noise \\
$T,\, T_e,\, T_d$ & Total / edge / device denoising steps \\
$s,\, s'$ & Relay step (edge) and start step (device) \\
$\sigma_t$ & Noise level at step $t$ \\
$\bar{\alpha}_t$ & Cumulative noise schedule coefficient \\
$\epsilon_\theta,\, v_\theta$ & Noise prediction / velocity field network \\
$\mathbf{e}$ & Prompt embeddings \\
$\delta_t$ & Per-step divergence between $\mathcal{M}_L$ and $\mathcal{M}_S$ \\
$\mathcal{M}_L,\, \mathcal{M}_S$ & Large (edge) and small (device) model \\
$\mathcal{A},\, a_t$ & Action space and selected arm at round $t$ \\
$\mathbf{c}_t,\, r_t$ & Context vector and reward at round $t$ \\
$Q_{\text{CLIP}},\, Q_{\text{IR}},\, Q_{\text{Pick}}$ & CLIP Score, ImageReward, PickScore \\
$Q_{\text{Aes}},\, Q_{\text{OCR}}$ & Aesthetic Score and OCR accuracy \\
\bottomrule
\end{tabular}
\end{table}

\subsection{Empirical Motivation}

The denoising process of diffusion models has a notable two-phase structure. In the early steps, when the signal-to-noise ratio $\mathrm{SNR}(t) = \bar{\alpha}_t / (1 - \bar{\alpha}_t)$ is low, the model establishes coarse semantic layout such as object placement and spatial composition. In the later steps, as the SNR increases, each step only adds fine-grained details that are nearly imperceptible to human viewers~\cite{choi2022p2}. This transition is also reflected in the attention mechanism: cross-attention outputs converge to a fixed point within the first few steps~\cite{liu2024tgate}, meaning that the generation trajectory becomes largely determined early on. These properties suggest that deploying a large model for the full denoising process is unnecessary. Instead, $\mathcal{M}_L$ can handle the early semantic-critical steps, and a lightweight $\mathcal{M}_S$ can take over the remaining refinement with limited quality degradation in our empirical setting.

\begin{figure}[t]
    \centering
    \begin{subfigure}{0.48\columnwidth}
        \centering
        \includegraphics[width=\textwidth]{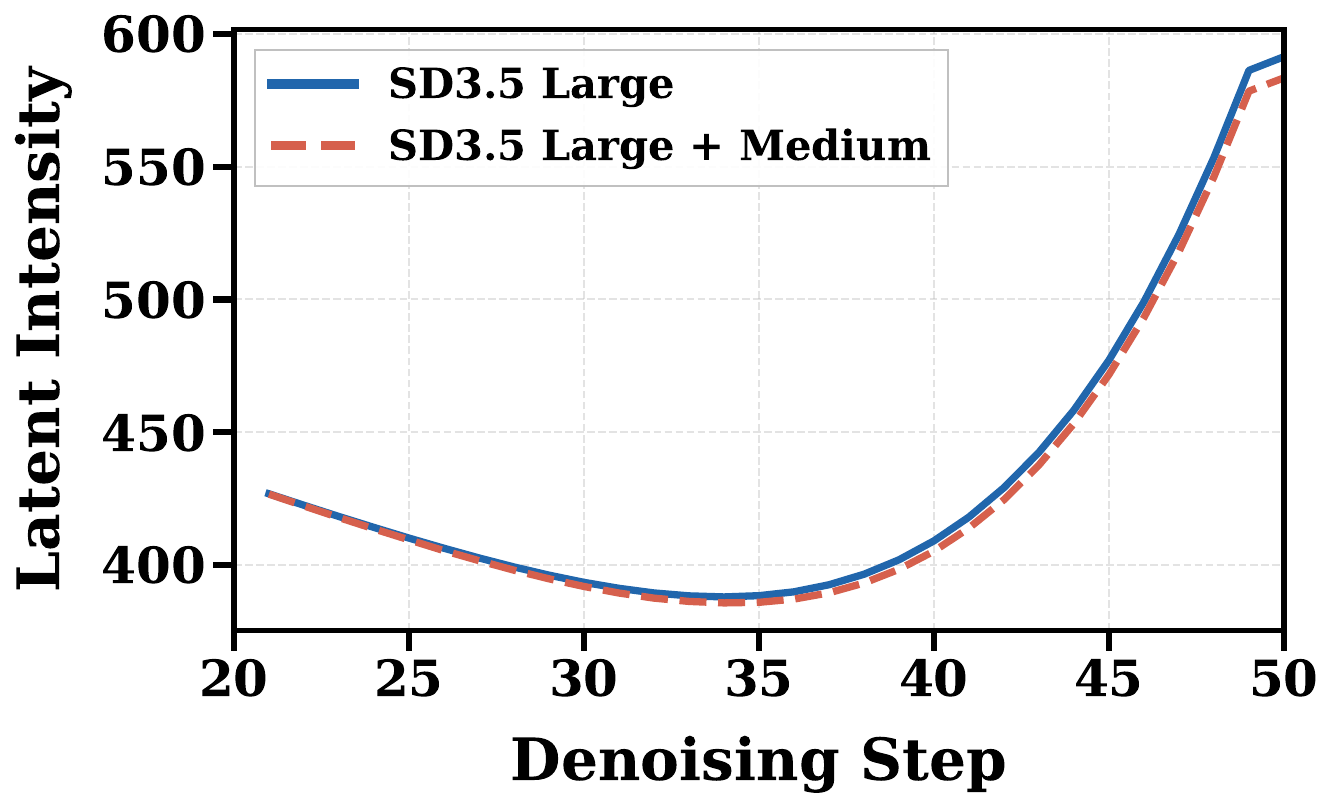}
        \caption{Latent intensity comparison}
        \label{fig:noise_intensity}
    \end{subfigure}%
    \hfill
    \begin{subfigure}{0.48\columnwidth}
        \centering
        \includegraphics[width=\textwidth]{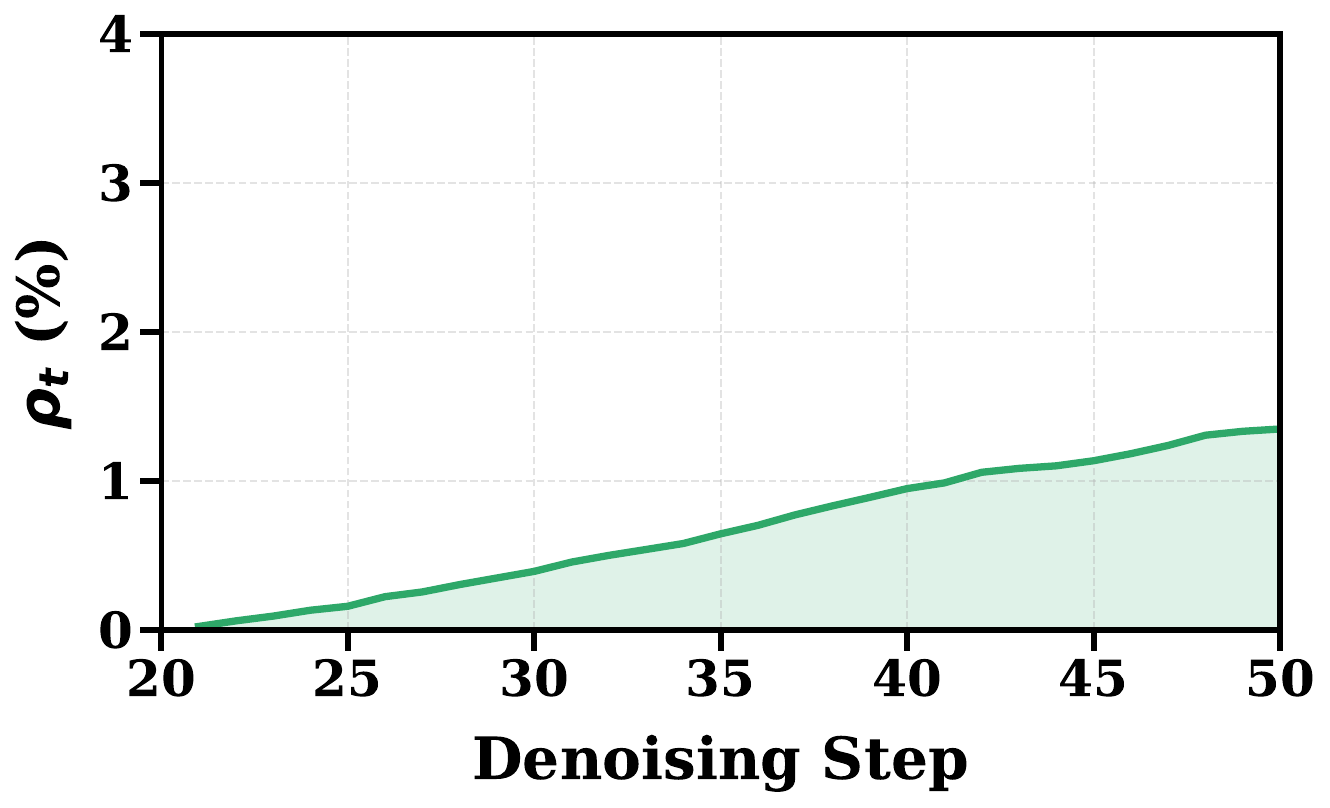}
        \caption{Per-step relative deviation $\rho_t$}
        \label{fig:latent_diff}
    \end{subfigure}
    \caption{Latent trajectory analysis}
    \label{fig:latent_analysis}
\end{figure}
We ran a latent trajectory comparison experiment on the SD3.5 family to test this idea. We let SD3.5 Large run the full 50 denoising steps, and also ran a relay setup where Large handled the first 20 steps and Medium took over for the remaining 30. After the relay point, we recorded the latent intensity $||\mathbf{x}_t||_2$ at each step. As Figure~\ref{fig:noise_intensity} shows, the two curves almost overlap after the handoff, meaning Medium did not deviate from the denoising direction that Large had established. To put a number on this gap, we calculated a per-step relative deviation:
\begin{equation}
    \rho_t = \frac{\left|\,\|\mathbf{x}_t^{\mathrm{large}}\|_2 -                                                     
  \|\mathbf{x}_t^{\mathrm{relay}}\|_2\,\right|}{\|\mathbf{x}_t^{\mathrm{large}}\|_2} \times 100\% 
\end{equation}

Figure~\ref{fig:latent_diff} shows that $\rho_t$ stays below 1.5\% throughout the relay phase with only minor fluctuations, which means the two models show similar latent-norm trajectories during the later denoising steps. These results show that, because models in the same family share the same latent space, the small model can naturally pick up where the large model left off.

\begin{figure*}[t]
    \centering
    \includegraphics[width=0.85\textwidth]{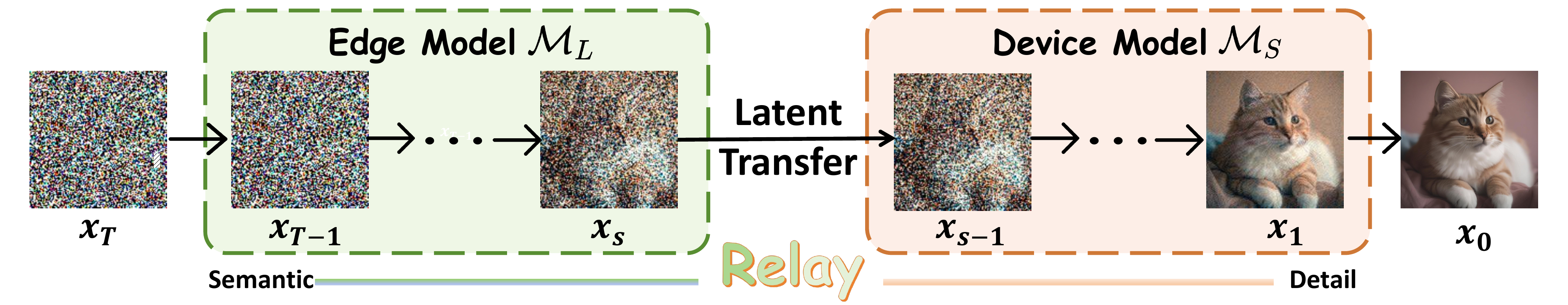}
    \caption{The process of relay inference}
    \label{fig:relay}
\end{figure*}

\subsection{Relay Implementation}

\textbf{Diffusion models.} Relay inference splits the denoising process at step~$s$. The large edge model $\mathcal{M}_L$ denoises the initial steps and produces an intermediate latent $x_s$. This latent is then transferred to the device where the small model $\mathcal{M}_S$ completes the generation. We show this overall process in Figure~\ref{fig:relay}. The exact step update depends on the model architecture. Models sharing a UNet backbone like the SDXL family typically follow the DDIM sampling framework~\cite{song2021ddim}, where each denoising step updates the latent as shown in Eq. (\ref{eq:ddim}):
\begin{equation}
    x_{t-1} = \sqrt{\bar{\alpha}_{t-1}}\,\hat{x}_0(x_t)
             + \sqrt{1-\bar{\alpha}_{t-1}}\;\epsilon_\theta(x_t, t),
    \label{eq:ddim}
\end{equation}
with $\hat{x}_0(x_t) = ({x_t - \sqrt{1{-}\bar{\alpha}_t}\;\epsilon_\theta(x_t,t)})/{\sqrt{\bar{\alpha}_t}}$. Here, $x_t$ represents the noisy latent at step $t$, $\bar{\alpha}_t$ is the cumulative noise schedule coefficient, $\epsilon_\theta$ denotes the noise prediction network, and $\hat{x}_0(x_t)$ is the estimated clean image. Models utilizing the MMDiT architecture like the SD3.5 family follow the Rectified Flow framework~\cite{liu2023rectifiedflow}. These models perform Euler integration for each step as shown in Eq. (\ref{eq:euler_step}):
\begin{equation}
    x_{i+1} = x_i + (t_{i+1} - t_i) \cdot v_\theta(x_i, t_i, \mathbf{e}).
    \label{eq:euler_step}
\end{equation}

At the relay point, the inference process switches from the edge model to the device model. The fundamental requirement for a successful handoff is maintaining noise continuity. We achieve this universal alignment through sigma matching, a strategy originally used in ensemble-of-experts diffusion pipelines~\cite{balaji2022ediffi}. Let $\{\sigma_i^{(e)}\}$ and $\{\sigma_j^{(d)}\}$ be the sigma sequences of $\mathcal{M}_L$ and $\mathcal{M}_S$. When $\mathcal{M}_L$ finishes step~$s$ with noise level $\sigma_s^{(e)}$, we calculate the device-side start step $s'$ by finding the closest noise level as shown in Eq. (\ref{eq:sigma_match}):
\begin{equation}
    s' = \arg\min_{j \in \{0,\dots,T_d-1\}} \left|\, \sigma_j^{(d)} - \sigma_s^{(e)} \,\right|.
    \label{eq:sigma_match}
\end{equation}

\textbf{Schedule compatibility.} The actual execution of this sigma matching theory depends on the default configurations and schedule compatibility within each model family. For the SDXL family, we set the total steps to $T_e{=}50$ for the edge model and $T_d{=}25$ for the device model. These models use different step counts and non-uniform Karras sigma schedules, which causes the same step index to map to different noise states. We therefore actively rely on the sigma matching equation and exhaustive search to find valid relay pairs. In contrast, both models in the SD3.5 family run the same $T{=}50$ steps on an identical linear schedule. This perfect alignment means the sigma matching trivially resolves to $s' = s$, allowing the small model to directly resume from step~$s$ at the same noise state. 

To cover different quality and latency tradeoffs, we select five relay points $s \in \{5, 10, 15, 20, 25\}$ for evaluation across both families. Instead of treating every denoising step as a separate relay action, we discretize the handoff point with a stride of five steps. This design reflects a service-level tradeoff: adjacent relay steps usually produce highly correlated latent states and similar quality--latency profiles, while a per-step action space would substantially increase the number of arms and dilute the feedback available to each arm in online bandit learning. This candidate set therefore provides a practical handoff granularity to capture the main quality--latency trend while keeping scheduling statistically stable under finite workloads.

The relay mechanism produces a set of configurations with different quality and speed characteristics within each family. In practice, the best choice depends on the request: a semantically complex prompt benefits from more edge-side steps to ensure coherence, while a simpler prompt can tolerate an earlier handoff for lower latency. User preferences and real-time factors such as device load and network conditions also affect which configuration performs best. Selecting the right configuration for each incoming request is therefore an online scheduling problem, which we formulate and solve in Section~\ref{sec:algorithm}.

\section{Algorithm Design}
\label{sec:algorithm}

While Section~\ref{sec:relay} defines the available relay configurations, selecting the optimal one for each incoming request remains non-trivial. The choice depends on dynamic factors like prompt complexity and system state, which are only known at runtime. Therefore, we formulate the scheduling problem as a contextual multi-armed bandit and propose a LinUCB-based online scheduling algorithm that learns the optimal context-to-configuration mapping through exploration--exploitation tradeoffs.

\subsection{Problem Formulation}

The relay mechanism produces a discrete set of inference configurations. We define an action space $\mathcal{A} = \{a_0, a_1, \dots, a_{10}\}$ of 11 arms, each mapping to one configuration shown in Table~\ref{tab:action_space}.

\begin{table}[t]
\centering
\caption{Action space of RISE.}
\label{tab:action_space}
\begin{tabular}{cll}
\toprule
\textbf{Arm} & \textbf{Model Pair} & \textbf{Relay Step} $s$ \\
\midrule
$a_0$             & Vega (standalone)        & --  \\
$a_1$--$a_5$      & SDXL + Vega              & 5, 10, 15, 20, 25 \\
$a_6$--$a_{10}$   & SD3.5 Large + Medium     & 5, 10, 15, 20, 25 \\
\bottomrule
\end{tabular}
\end{table}

Each request is encoded as a $d{=}8$ dimensional context vector:
\begin{equation}
    \mathbf{c} = [\, c_\mathrm{cplx},\; c_\mathrm{txt},\; c_\mathrm{net},\; c_\mathrm{bat},\; c_\mathrm{pref},\; l_\mathrm{vega},\; l_\mathrm{sdxl},\; l_\mathrm{sd3} \,].
    \label{eq:context}
\end{equation}

The first five dimensions capture \emph{task-level} features:
\begin{itemize}
    \item $c_\mathrm{cplx} \in [0,1]$: prompt complexity, measured as the normalized clause count.
    \item $c_\mathrm{txt} \in \{0,1\}$: text-rendering indicator, set to 1 when the prompt requests visible text.
    \item $c_\mathrm{net} \in [0,1]$: inverse network quality, derived from the log-transformed round-trip latency. Higher values favor device-side inference.
    \item $c_\mathrm{bat} \in \{0,1\}$: low-battery flag, set to 1 when the device battery drops below 20\%.
    \item $c_\mathrm{pref} \in [0,1]$: quality--speed preference, where 0 denotes maximum quality and 1 maximum speed.
\end{itemize}

The remaining three dimensions encode \emph{system-level} state:
\begin{itemize}
    \item $l_\mathrm{vega},\, l_\mathrm{sdxl},\, l_\mathrm{sd3} \in [0,1]$: occupancy ratios of the Vega, SDXL, and SD3.5 device pools, each equaling the fraction of occupied replicas. Relay arms occupy slots from two pools simultaneously, so LinUCB observes both loads when evaluating such arms, implicitly steering away from congested pools.
\end{itemize}

\textbf{Optimization objective.} At round $t$, a request arrives with context $\mathbf{c}_t$. The scheduler picks arm $a_t \in \mathcal{A}$ and observes reward $r_t$ (defined in Section~\ref{sec:reward}). Given $N$ total requests, the objective is to maximize cumulative reward:
\begin{equation}
    \max_{\{a_t\}_{t=1}^{N}} \sum_{t=1}^{N} r_t.
    \label{eq:objective}
\end{equation}

This problem has a natural exploration--exploitation tradeoff, which we address with a LinUCB-based approach~\cite{li2018linucb}.

\subsection{LinUCB Scheduler}

For each arm $a \in \mathcal{A}$, the algorithm maintains a matrix $\mathbf{A}_a \in \mathbb{R}^{d \times d}$ and a vector $\mathbf{b}_a \in \mathbb{R}^d$, initialized as $\mathbf{A}_a = \mathbf{I}_d$ and $\mathbf{b}_a = \mathbf{0}$, together with a pull count $n_a = 0$. The selection procedure has three steps.

\textbf{UCB scoring.} Given context $\mathbf{c}_t$, the score for each candidate arm has three components:
\begin{equation}
    p_a = \underbrace{\hat{\boldsymbol{\theta}}_a^\top \mathbf{c}_t}_{\text{exploitation}} + \underbrace{\alpha \sqrt{\mathbf{c}_t^\top \mathbf{A}_a^{-1} \mathbf{c}_t}}_{\text{contextual exploration}} + \underbrace{\beta \sqrt{\frac{\ln(n+1)}{1 + n_a}}}_{\text{frequency exploration}},
    \label{eq:ucb}
\end{equation}
where $\hat{\boldsymbol{\theta}}_a = \mathbf{A}_a^{-1}\mathbf{b}_a$ is the ridge regression estimate and $n = \sum_a n_a$ is the total pull count. The first term estimates expected reward from historical data; the second, following the standard LinUCB formulation~\cite{li2018linucb}, adds a bonus for contexts poorly covered in past observations; the third, inspired by UCB1~\cite{auer2002ucb}, provides additional exploration for less-pulled arms.

\textbf{Softmax sampling.} Instead of always selecting the arm with the highest score, we draw from a softmax distribution:
\begin{equation}
    \Pr(a_t = a) = \frac{\exp(p_a / \tau)}{\sum_{a' \in \mathcal{A}_t} \exp(p_{a'} / \tau)},
    \label{eq:softmax}
\end{equation}
where $\tau$ is a temperature parameter. Softmax sampling serves two purposes: in early rounds it maintains stochastic exploration when reward estimates are still uncertain, and in later rounds it prevents the scheduler from locking onto one arm and falling into a local optimum. The temperature decays over time:
\begin{equation}
    \tau = \max\!\left(\tau_{\min},\; \tau_0 \cdot \left(1 - \frac{\max(0,\, n - N_w)}{K}\right)\right),
    \label{9}
\end{equation}
where $\tau_0$ is the initial temperature, $N_w$ is the warm-up period, and $K$ is a shared decay constant. This schedule gradually shifts behavior from broad exploration to focused exploitation.

\textbf{Parameter update.} After selecting arm $a_t$ and observing reward $r_t$, the selected arm's statistics are updated:
\begin{align}
    \mathbf{A}_{a_t} &\leftarrow \mathbf{A}_{a_t} + \mathbf{c}_t \mathbf{c}_t^\top + \lambda \mathbf{I}_d, \label{eq:update} \\
    \mathbf{b}_{a_t} &\leftarrow \mathbf{b}_{a_t} + r_t \, \mathbf{c}_t, \nonumber
\end{align}
where $\lambda$ is a regularization coefficient. Unlike the standard LinUCB update, we add $\lambda \mathbf{I}_d$ at every step so that the shrinkage on $\hat{\boldsymbol{\theta}}_a$ grows with the number of pulls, preventing overfitting when the reward distribution shifts. The exploration parameters $\alpha$ and $\beta$ decay once the total pull count exceeds $N_w$:
\begin{equation}
    \alpha = \max\!\left(\alpha_{\min},\; \alpha_0 - \frac{\max(0,\, n - N_w)}{K}\right),
    \label{eq:alpha_decay}
\end{equation}
\begin{equation*}
    \beta = \max\!\left(\beta_{\min},\; \beta_0 \cdot \left(1 - \frac{\max(0,\, n - N_w)}{K}\right)\right).
\end{equation*}

Algorithm~\ref{alg:linucb} details the complete arm selection procedure.

\begin{algorithm}[t]
\caption{RISE Arm Selection}
\label{alg:linucb}
\begin{algorithmic}[1]
\REQUIRE Candidate arms $\mathcal{A}_t$, context $\mathbf{c}_t$, parameters $\{\mathbf{A}_a, \mathbf{b}_a, n_a\}$
\STATE $n \leftarrow \sum_{a} n_a$
\STATE Update $\alpha, \beta, \tau$ via Eq. (\ref{9}) and Eq. (\ref{eq:alpha_decay})
\FOR{each $a \in \mathcal{A}_t$}
    \STATE $\hat{\boldsymbol{\theta}}_a \leftarrow \mathbf{A}_a^{-1} \mathbf{b}_a$
    \STATE $p_a \leftarrow \hat{\boldsymbol{\theta}}_a^\top \mathbf{c}_t + \alpha \sqrt{\mathbf{c}_t^\top \mathbf{A}_a^{-1} \mathbf{c}_t} + \beta \sqrt{\frac{\ln(n{+}1)}{1 + n_a}}$
\ENDFOR
\RETURN Sample $a_t$ from $\mathcal{A}_t$ via softmax 
\end{algorithmic}
\end{algorithm}

\subsection{Dynamic Reward Function}
\label{sec:reward}

The reward signal guides RISE toward the best-matching configuration for each task. We define a composite reward:
\begin{equation}
    r = \!\sum_{m \in \mathcal{Q}} w_m Q_m - w_\mathrm{time} \cdot t_\mathrm{total} - w_\mathrm{cost} \cdot m_\mathrm{vram} - \gamma \cdot l_\mathrm{dev},
    \label{eq:reward_raw}
\end{equation}
where $\mathcal{Q} = \{Q_\text{CLIP},\, Q_\text{IR},\, Q_\text{Pick},\, Q_\text{Aes},\, Q_\text{OCR}\}$ represents five quality metrics. The cost consists of three components: the end-to-end latency $t_\mathrm{total}$ including queuing time, the peak GPU memory $m_\mathrm{vram}$, and the occupancy ratio $l_\mathrm{dev}$ of the device pools. For relay arms spanning two distinct pools, $l_\mathrm{dev}$ is defined as the maximum of the two ratios. All cost terms are collectively weighted by a penalty coefficient $\gamma$.

Dynamic weights empower metric differences across configurations to  govern arm selection:
\begin{itemize}
    \item \emph{Text-rendering tasks} ($c_\mathrm{txt}{=}1$): $w_\mathrm{OCR}$ is raised while visual quality weights drop.
    \item \emph{Speed-sensitive tasks} ($c_\mathrm{pref}{>}0.5$): $w_\mathrm{time}$ is amplified and quality weights are halved.
    \item \emph{Quality-focused tasks}: $w_\mathrm{CLIP}$ and $w_\mathrm{IR}$ are maximized and $w_\mathrm{time}$ is reduced.
    \item \emph{Low battery} ($c_\mathrm{bat}{=}1$): both $w_\mathrm{cost}$ and $w_\mathrm{time}$ are scaled up to favor energy-efficient setups.
\end{itemize}

To keep the reward bounded and stabilize learning, we compress it through a $\tanh$ function:
\begin{equation}
    r_\mathrm{final} = \eta \cdot \tanh(r / \eta),
    \label{eq:reward_final}
\end{equation}
where $\eta$ is a scaling constant that keeps $r_\mathrm{final} \in (-\eta, \eta)$. Throughout this paper, $r_t$ denotes the compressed reward.

\begin{algorithm}[t]
\caption{RISE Online Scheduling}
\label{alg:rise}
\begin{algorithmic}[1]
\REQUIRE Action space $\mathcal{A}$, minimum pulls $N_\mathrm{min}$
\STATE Initialize $\mathbf{A}_a \leftarrow \mathbf{I}_d,\; \mathbf{b}_a \leftarrow \mathbf{0},\; n_a \leftarrow 0$ for all $a \in \mathcal{A}$
\FOR{each request $t$}
    \STATE Construct context $\mathbf{c}_t$ via Eq. (\ref{eq:context})
    \STATE Filter $\mathcal{A}_t \subseteq \mathcal{A}$ by device availability
    \IF{$\mathcal{A}_t = \emptyset$}
        \STATE Enqueue request; \textbf{continue}
    \ENDIF
    \IF{$\exists\, a \in \mathcal{A}_t$ with $n_a < N_\mathrm{min}$}
        \STATE $a_t \leftarrow \arg\min_{a \in \mathcal{A}_t} n_a$ \COMMENT{Forced exploration}
    \ELSE
        \STATE $a_t \leftarrow$ \textsc{LinUCBArmSelect}($\mathcal{A}_t$, $\mathbf{c}_t$) \COMMENT{Alg.~\ref{alg:linucb}}
    \ENDIF
    \STATE Execute relay inference with $a_t$
    \STATE Compute $r_t$ via Eqs. \eqref{eq:reward_raw}--\eqref{eq:reward_final}
    \STATE Update $\mathbf{A}_{a_t},\, \mathbf{b}_{a_t},\, n_{a_t}$ using Eq. (\ref{eq:update})
\ENDFOR
\end{algorithmic}
\end{algorithm}

The regret property of standard LinUCB motivates our use of confidence-based exploration in RISE. However, since the practical scheduler further incorporates softmax sampling, frequency-based exploration, and workload-aware reward shaping, we do not claim a formal regret bound for the complete scheduler. 

Combining all components, the system constructs a context for each arriving request, filters available arms, selects a configuration via LinUCB, executes inference, and updates parameters based on the observed reward. Algorithm~\ref{alg:rise} presents the complete RISE online scheduling procedure. Notably, $t_\mathrm{total}$ includes queuing wait time, so RISE naturally learns to avoid choosing an arm on a busy pool, which leads to longer waits and lower rewards, discouraging that choice in similar future contexts.

\section{Experiment}
\label{sec:experiment}

We design experiments to answer three research questions:
\begin{itemize}
    \item \textbf{RQ1:} Can relay inference achieve significant speedup while preserving generation quality across heterogeneous tasks?
    \item \textbf{RQ2:} Does the RISE scheduler effectively balance quality and latency under mixed service workloads?
    \item \textbf{RQ3:} How does each component of the scheduling algorithm contribute to the overall service performance?
\end{itemize}

\subsection{Experimental Setup}

\textbf{Datasets.} We evaluate RISE on two datasets covering complementary aspects of text-to-image generation services. \emph{DiffusionDB}~\cite{wang2022diffusiondb} is a large-scale prompt gallery dataset covering diverse visual scenes; \emph{DrawTextCreative}~\cite{chen2023textdiffuser} contains prompts requiring legible text within the image, testing typographic fidelity alongside visual quality.

\begin{table*}[t]
\renewcommand{\arraystretch}{1.3}
\caption{Performance comparison of generation quality and efficiency across different acceleration methods on two datasets. OCR is omitted for DiffusionDB as its prompts lack text-rendering tasks. And the best results per group are \textbf{bolded} throughout this paper.}
\label{table_comparison}
\centering
\scriptsize
\newcommand{\gc}{\cellcolor{gray!15}} % 定义一个简化的命令，仅用于需要加灰底的单元格
\resizebox{\textwidth}{!}{
\begin{tabular}{ccc|ccccc|cc}
\toprule
\multirow{2}{*}{\textbf{Dataset}} & \multirow{2}{*}{\textbf{Model}} & \multirow{2}{*}{\textbf{Method}} & \multicolumn{5}{c|}{\textbf{Quality Metrics}} & \multicolumn{2}{c}{\textbf{Service Efficiency}} \\ \cmidrule{4-10}
 & & & \textbf{CLIP}$\uparrow$ & \textbf{ImgRwd}$\uparrow$ & \textbf{PickSc}$\uparrow$ & \textbf{Aesth}$\uparrow$ & \textbf{OCR}$\uparrow$ & \textbf{Speedup}$\uparrow$ &
\textbf{Denoise(s)}$\downarrow$ \\ \midrule

% =============== DiffusionDB ===============
 & & Original & 0.3385 & 0.8482 & 0.2150 & 6.4453 & -- & 1$\times$ & 6.87 \\
 & & DeepCache & 0.3255 & 0.7559 & 0.2145 & 6.4576 & -- & 1.85$\times$ & 3.72 \\ 
 & & T-GATE & 0.3385 & 0.5728 & 0.2124 & 6.4702 & -- & 1.45$\times$ & 4.54 \\
 & & SADA & 0.3387 & 0.5847  & 0.2141 & 6.4258 & -- & 1.87$\times$ & 3.67 \\
 & & \gc RISE (Fast)& \gc 0.3392 & \gc 1.0598 & \gc \textbf{0.2153} & \gc 6.4969 & \gc -- & \gc \textbf{2.10}$\times$ & \gc \textbf{3.27} \\ 
 & \multirow{-6}{*}{SDXL} & \gc RISE (Slow)& \gc \textbf{0.3397} & \gc \textbf{1.0901} & \gc 0.2150 & \gc \textbf{6.5104} & \gc -- & \gc 1.59$\times$ & \gc 4.33\\ \cmidrule{2-10}

 & & Original & 0.3328 & 1.1289 & 0.2188 & 6.4164 & -- & 1$\times$ & 30.19 \\
 & & DeepCache & 0.3320 & 1.0880 & 0.2186 & 6.4283 & -- & 1.30$\times$ & 23.16\\
 & & T-GATE & 0.3185 & 0.6612 & 0.1983 & 6.2597 & -- & 1.62$\times$ & 18.60 \\
 & & SADA & 0.3278 & 0.9645  & 0.2064 & 6.3122 & -- & 1.32$\times$ & 22.83 \\
 & & \gc RISE (Fast)& \gc \textbf{0.3335} & \gc 1.1209 & \gc 0.2195 & \gc 6.4411 & \gc -- & \gc \textbf{1.77}$\times$ & \gc \textbf{17.10} \\ 
 \multirow{-12}{*}{\textbf{DiffusionDB}} & \multirow{-6}{*}{SD3.5 Large} & \gc RISE (Slow)& \gc 0.3329 & \gc \textbf{1.1303} & \gc \textbf{0.2198} & \gc \textbf{6.4417} & \gc -- & \gc 1.59$\times$ & \gc 18.95\\
\midrule

% =============== DrawText-Creative ===============
 & & Original& 0.3300 & 0.6521 & 0.2179 & 5.1366 & 0.2551 & 1$\times$ & 6.89 \\
 & & DeepCache & 0.3220 & 0.3603 & \textbf{0.2129} & 5.1606 & \textbf{0.1879} &1.84$\times$ & 3.75 \\ 
 & & T-GATE & 0.3180 & 0.0117 & 0.2076 & \textbf{5.1705} & 0.1126 & 1.53$\times$ & 4.49 \\
 & & SADA & 0.3220 & \textbf{0.4422}  & 0.2033 & 5.0281 & 0.1196 & 1.87$\times$ & 3.67 \\
 & & \gc RISE (Fast)& \gc \textbf{0.3264} & \gc 0.2932 & \gc 0.2075 & \gc 5.0371 & \gc 0.0956 & \gc \textbf{2.10}$\times$ & \gc \textbf{3.28} \\ 
 & \multirow{-6}{*}{SDXL} & \gc RISE (Slow)& \gc 0.3236 & \gc 0.2402 & \gc 0.2082 & \gc 5.0146 & \gc 0.1016 & \gc 1.58$\times$ & \gc 4.35\\ \cmidrule{2-10}

 & & Original & 0.3583 & 1.2855 & 0.2298 & 5.0888 & 0.6638 & 1$\times$ & 30.92 \\
 & & DeepCache & 0.3573 & 1.2341 & 0.2291 & 5.0336 & 0.6132 & 1.33$\times$ & 23.27\\
  & & T-GATE & 0.3420 & 0.6123 & 0.2140 & 4.6719 & 0.5746 & 1.66$\times$ & 18.61 \\
  & & SADA & 0.3421 & 1.1964  & 0.2196 & 4.9824 & 0.6254 & 1.34$\times$ & 23.05 \\
 & & \gc RISE (Fast)& \gc \textbf{0.3583} & \gc 1.2670 & \gc 0.2298 & \gc 5.0768 & \gc 0.6695 & \gc \textbf{1.77}$\times$ & \gc \textbf{17.47} \\ 
 \multirow{-12}{*}{\textbf{\makecell{DrawText-\\Creative}}} & \multirow{-6}{*}{SD3.5 Large} & \gc RISE (Slow)& \gc 0.3580 & \gc \textbf{1.3052} & \gc \textbf{0.2300} & \gc \textbf{5.0954} & \gc \textbf{0.6771} & \gc 1.60$\times$ & \gc 19.32\\
\bottomrule

\end{tabular}
}
\end{table*}

\textbf{Models and Testbed.} We employ two model families for relay inference. For the SDXL family, the edge model $\mathcal{M}_L$ is SDXL and the device model $\mathcal{M}_S$ is Segmind-Vega, sharing a UNet backbone with a 4-channel latent space. For the SD3 family, $\mathcal{M}_L$ is Stable Diffusion 3.5 Large and $\mathcal{M}_S$ is SD3.5 Medium, sharing the MMDiT architecture with a 16-channel latent space. Our testbed comprises 8 NVIDIA RTX 4090 GPUs organized as 4 device pools (SDXL$\times$2, SD3.5 Large$\times$2, SD3.5 Medium$\times$2, Segmind-Vega$\times$2). Although network and energy factors are simulated in the scheduling context, the testbed does not include deployment on physical heterogeneous edge-device hardware. All models generate images at $1024 \times 1024$ resolution. SDXL uses $T_e{=}50$ steps and Segmind-Vega uses $T_d{=}25$ steps; SD3.5 Large and Medium both use $T{=}50$ steps.

\textbf{Evaluation Metrics.} We assess generation quality with five metrics: CLIP Score~\cite{radford2021clip} for text--image semantic alignment, ImageReward~\cite{xu2023imagereward} for human preference prediction, PickScore~\cite{kirstain2023pickscore} for pairwise preference alignment, Aesthetic Score~\cite{schuhmann2022laion} for visual appeal, and OCR accuracy for text-rendering fidelity on DrawTextCreative, where we use ABINet~\cite{fang2021abinet} to recognize text in generated images and compute the normalized edit distance. Service efficiency is measured by denoising latency and the corresponding speedup ratio relative to each family's full model.

\textbf{Baselines.} We adopt two groups of baselines to evaluate our approach.
(1)~\emph{Inference acceleration baselines}: for both the SDXL and SD3.5 families, we compare against the original full models (SDXL and SD3.5 Large) running all 50 steps independently. To further demonstrate the effectiveness of our approach, we also include training-free acceleration methods applied to the large models:
\begin{itemize}
    \item \textbf{DeepCache}~\cite{ma2024deepcache}: caches high-level features across adjacent denoising steps to reduce redundant computations.
    \item \textbf{T-GATE}~\cite{liu2025faster}: skips cross-attention computations after semantic content converges.
    \item \textbf{SADA}~\cite{jiang2025sada}: a stability-guided adaptive diffusion acceleration method that dynamically allocates sparsity.
\end{itemize}
(2)~\emph{Scheduling baselines}: these methods focus on how to dispatch requests and select configurations online:
\begin{itemize}
    \item \textbf{Round-Robin (RR)}: cycles through available configurations in a fixed, sequential order.
    \item \textbf{Greedy}: a makespan-minimizing heuristic that dispatches to the least-loaded pool with a fixed mid-range relay step.
    \item \textbf{PPO}~\cite{schulmanproximal}: a policy-gradient reinforcement learning method for stable policy updates.
    \item \textbf{SAC}~\cite{haarnoja2018sac}: an off-policy actor-critic method that incorporates entropy regularization.
\end{itemize}

\subsection{Relay Inference Performance (RQ1)}

Table~\ref{table_comparison} presents the quantitative comparison across both datasets. Following the action space in Table~\ref{tab:action_space}, we select two representative relay configurations from each family: one optimized for speed (Fast, $s=15$) and one for quality (Slow, $s=20$). For example, the `Fast' method for SDXL model denotes SDXL+Segmind-Vega relay with an early handoff step ($s=15$), and the `Slow' method for SD3.5 Large model means SD
3.5 Large+SD3.5 Medium relay with a later handoff step ($s=20$). Baseline settings include DeepCache (\texttt{cache interval=2}) and T-GATE (\texttt{gate\_step=20}). Speedup is relative to the original full model. Moreover, we summarize two key findings.

\textbf{Finding 1: Relay inference achieves significant speedups and superior metric robustness compared to both full models and single-model acceleration baselines.} 
On the DiffusionDB dataset, relay configurations outperform not only their full-model counterparts but also training-free acceleration baselines in human preference and aesthetic metrics. While existing single-model baselines effectively reduce latency, they often do so at a severe cost to semantic alignment and structural integrity. In contrast, by executing the crucial early denoising steps natively on the large edge model and delegating detail refinement to the small device model, RISE maintains semantic coherence. This complementary approach delivers highly competitive speedups without sacrificing generation quality, offering a more favorable quality-latency tradeoff.

\begin{figure*}[t]
    \centering
    \vspace{1mm}
    \newcommand{\cw}{0.13\textwidth}
    \setlength{\tabcolsep}{1pt}
    \renewcommand{\arraystretch}{0}
    \begin{tabular}{@{}m{\cw} m{\cw} m{\cw} m{0.12\textwidth} m{\cw} m{\cw} m{\cw}@{}}
    % --- Column headers ---
    \centering\fontsize{7}{8}\selectfont \textbf{SDXL} &
    \centering\fontsize{7}{8}\selectfont \textbf{RISE (Fast)} &
    \centering\fontsize{7}{8}\selectfont \textbf{RISE (Slow)} &
    \centering\fontsize{7}{8}\selectfont \textbf{Prompt} &
    \centering\fontsize{7}{8}\selectfont \textbf{SD3.5 Lar.} &
    \centering\fontsize{7}{8}\selectfont \textbf{SD3.5 Med.} &
    \centering\arraybackslash\fontsize{7}{8}\selectfont \textbf{RISE (Slow)} \\
    % --- Speedup row ---
    \centering\fontsize{7}{8}\selectfont 1$\times$ &
    \centering\fontsize{7}{8}\selectfont 2.10$\times$ &
    \centering\fontsize{7}{8}\selectfont 1.59$\times$ &
    &
    \centering\fontsize{7}{8}\selectfont 1$\times$ &
    \centering\fontsize{7}{8}\selectfont 2.67$\times$ &
    \centering\arraybackslash\fontsize{7}{8}\selectfont 1.59$\times$ \\[1pt]
    % --- Row 1 ---
    \includegraphics[width=\cw]{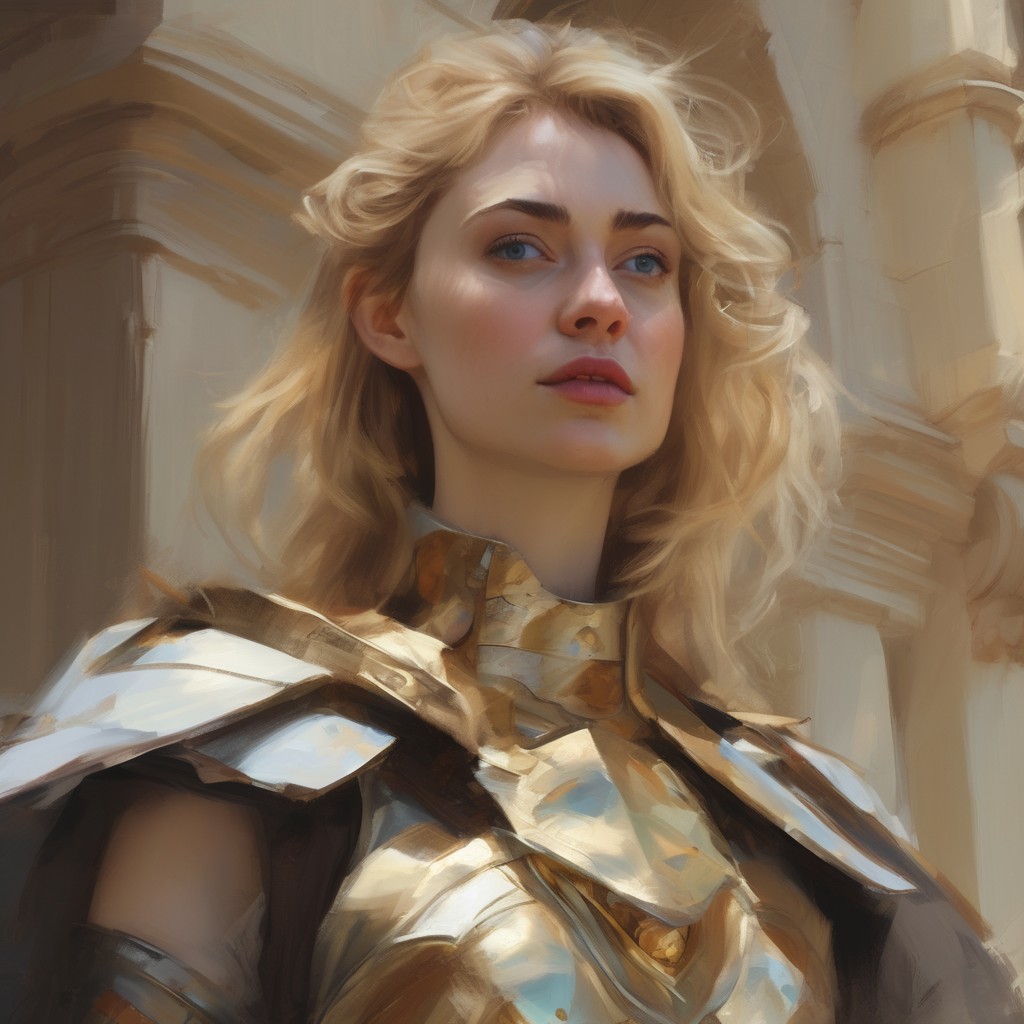} &
    \includegraphics[width=\cw]{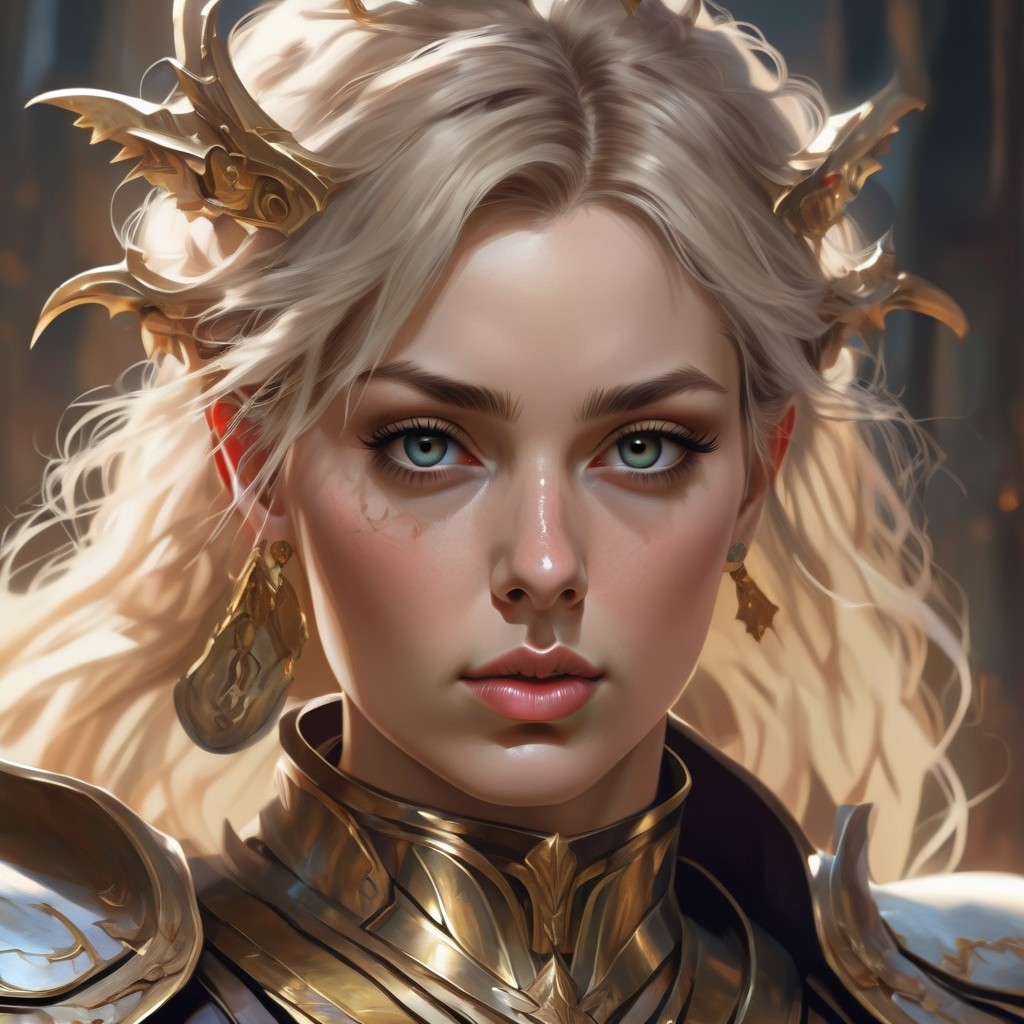} &
    \includegraphics[width=\cw]{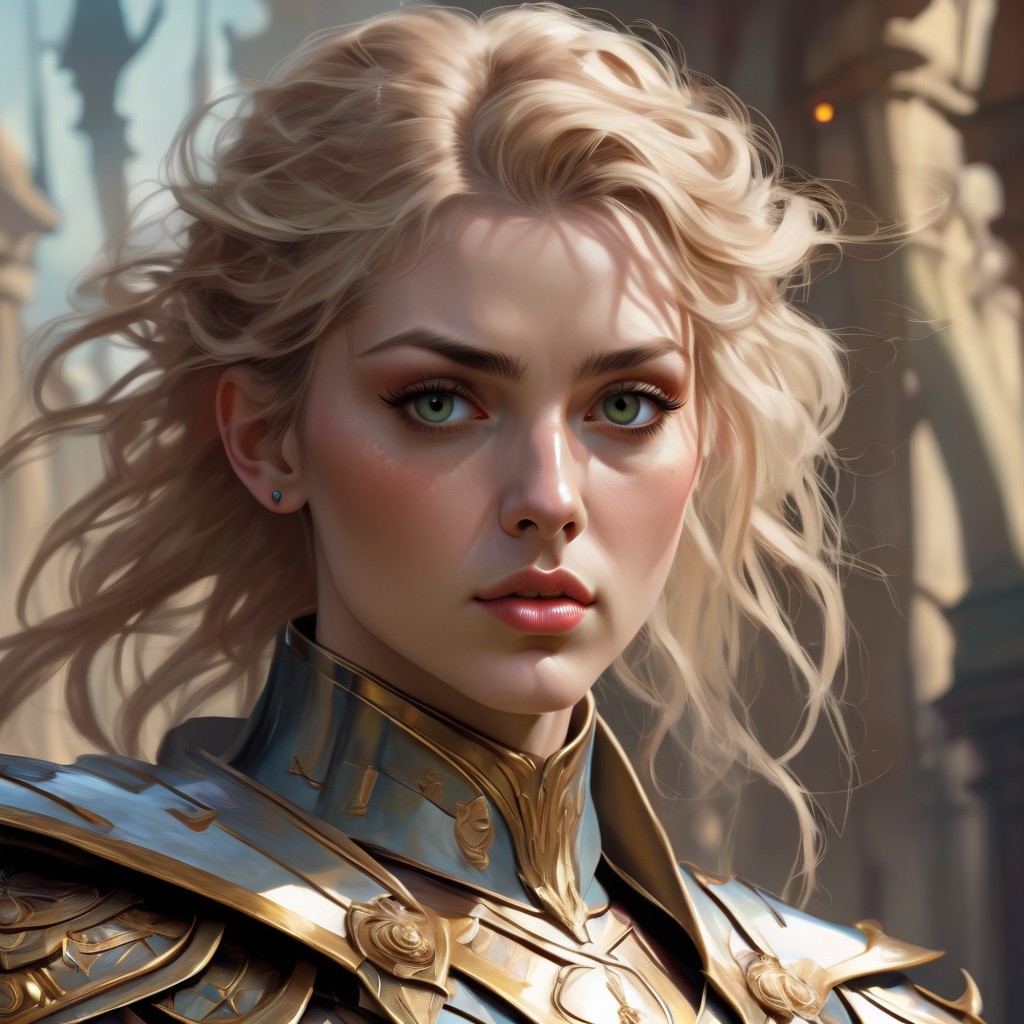} &
    \parbox[c]{0.12\textwidth}{\raggedright\fontsize{7}{8}\selectfont dynamic dramatic portrait, imogen poots as battle cleric goddess, ...} &
    \includegraphics[width=\cw]{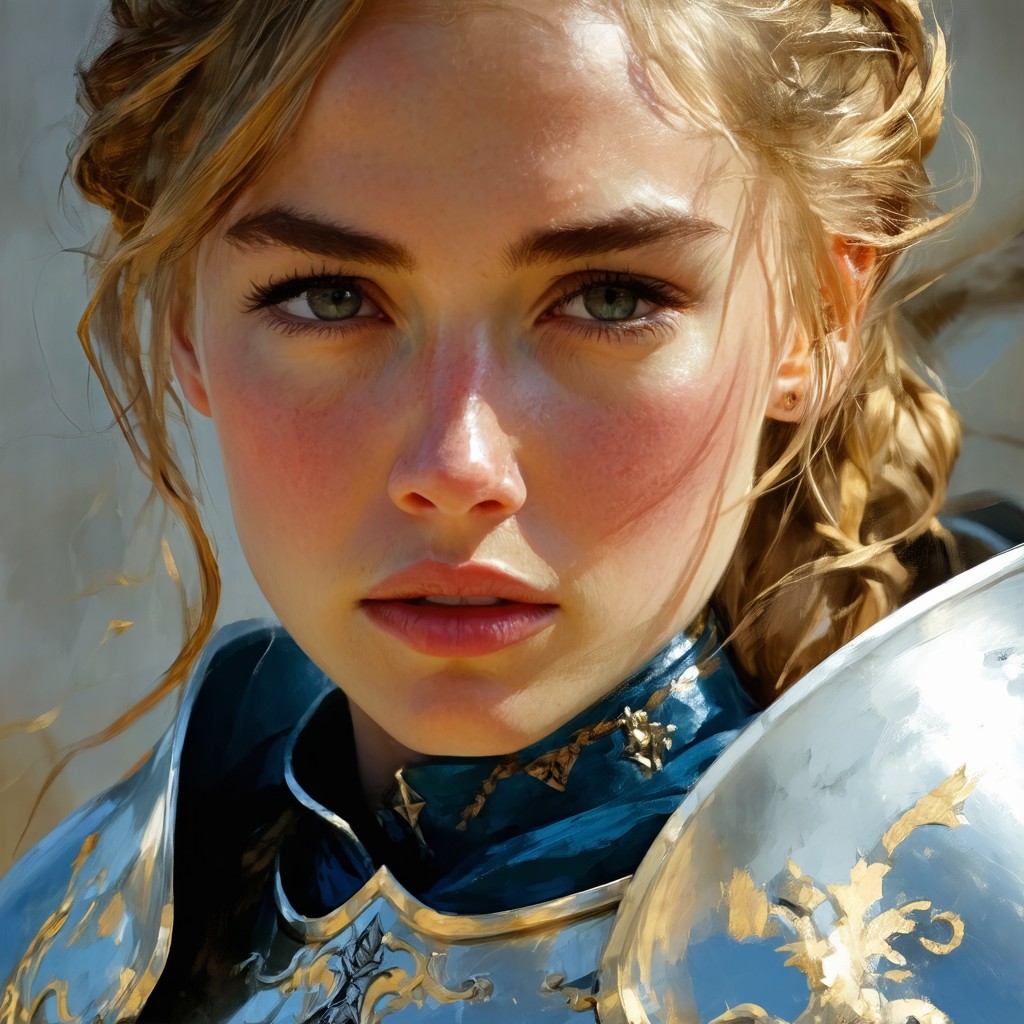} &
    \includegraphics[width=\cw]{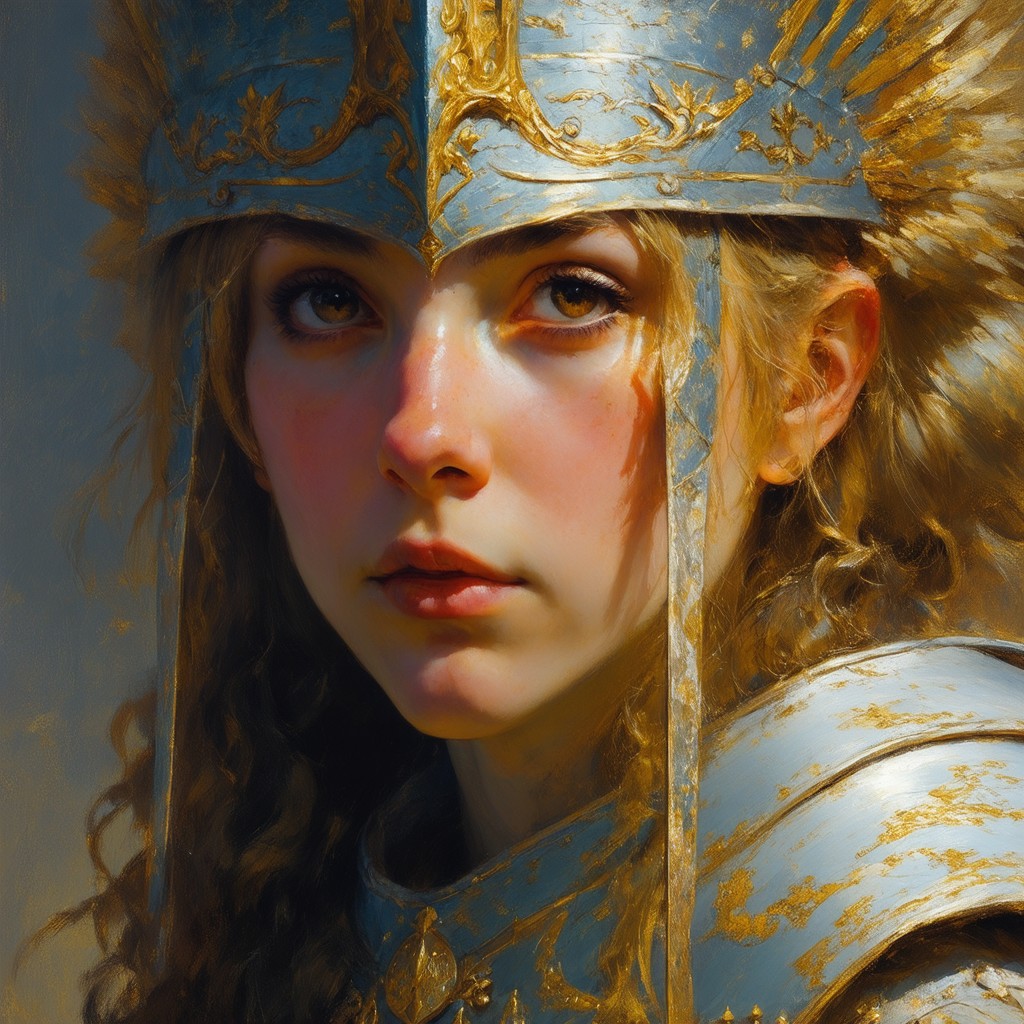} &
    \includegraphics[width=\cw]{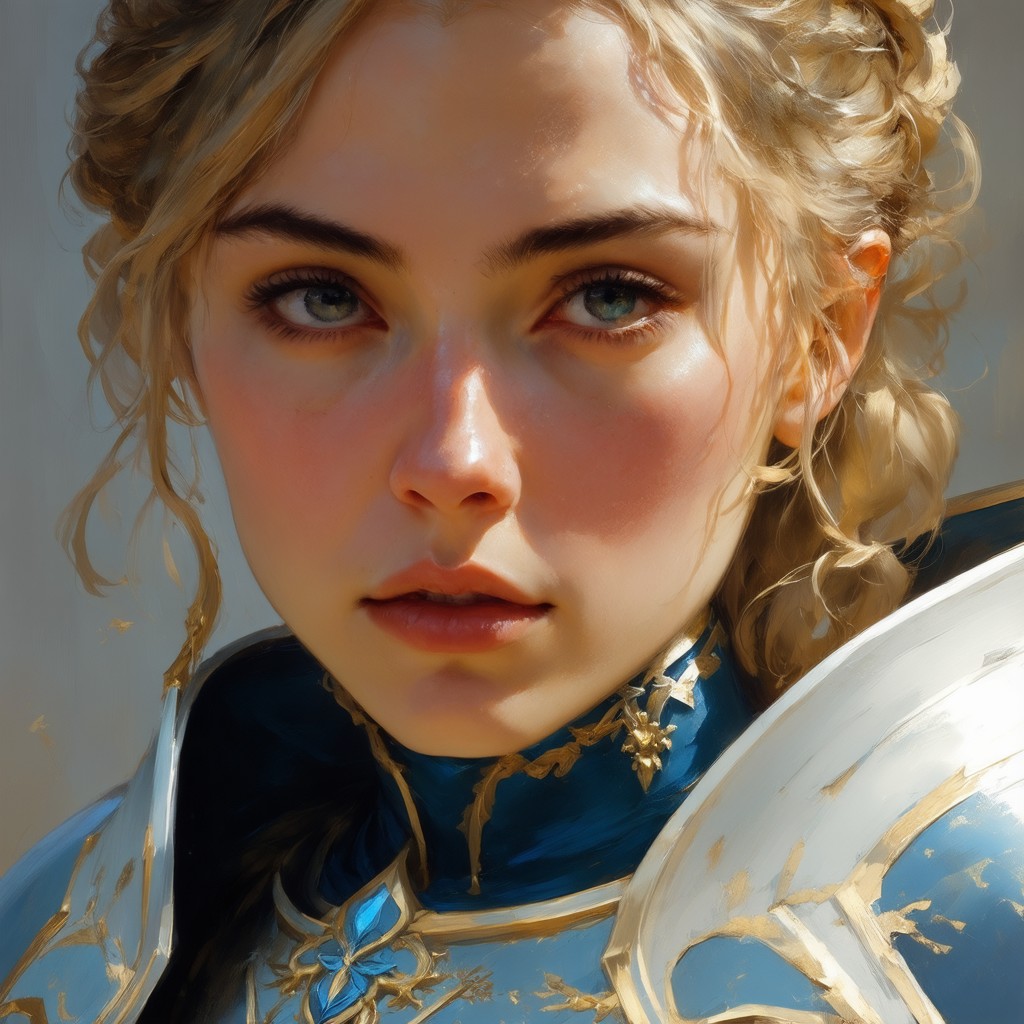} \\[1pt]
    % --- Row 2 ---
    \includegraphics[width=\cw]{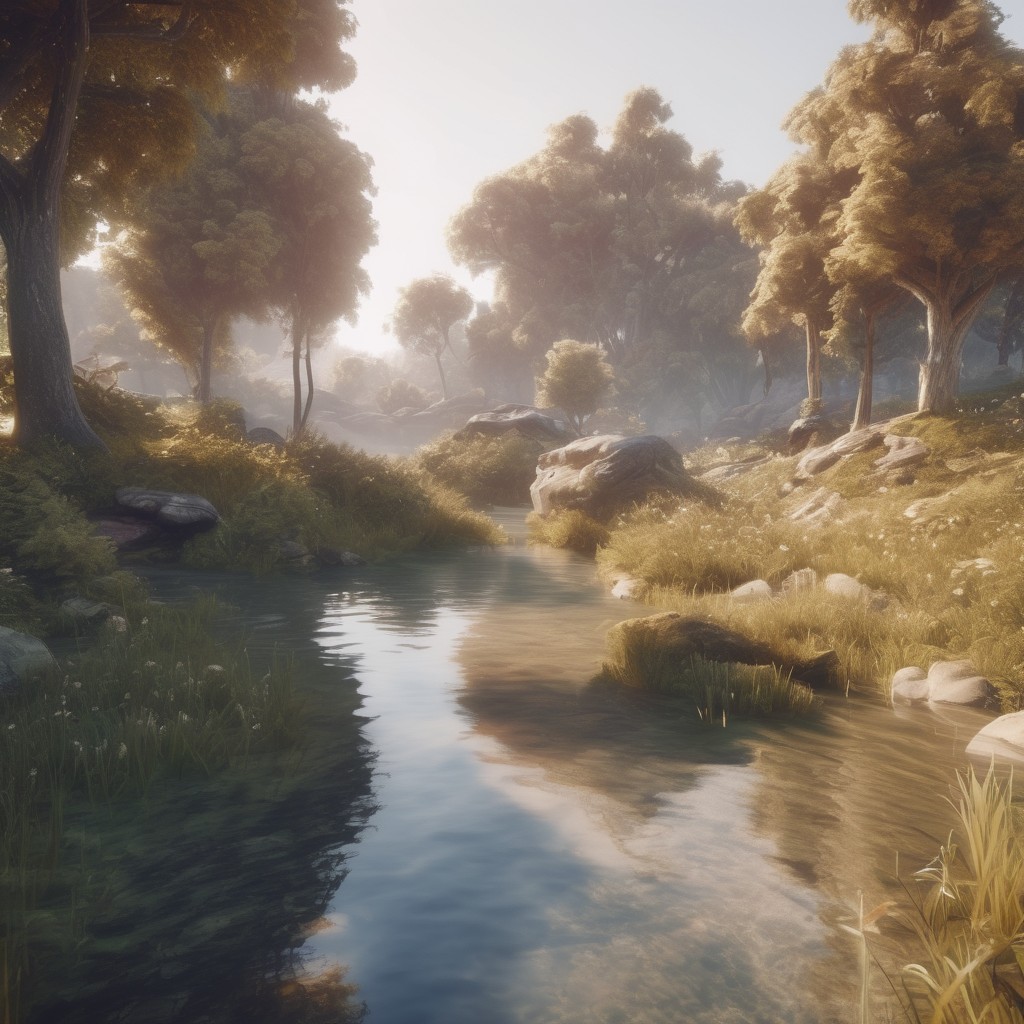} &
    \includegraphics[width=\cw]{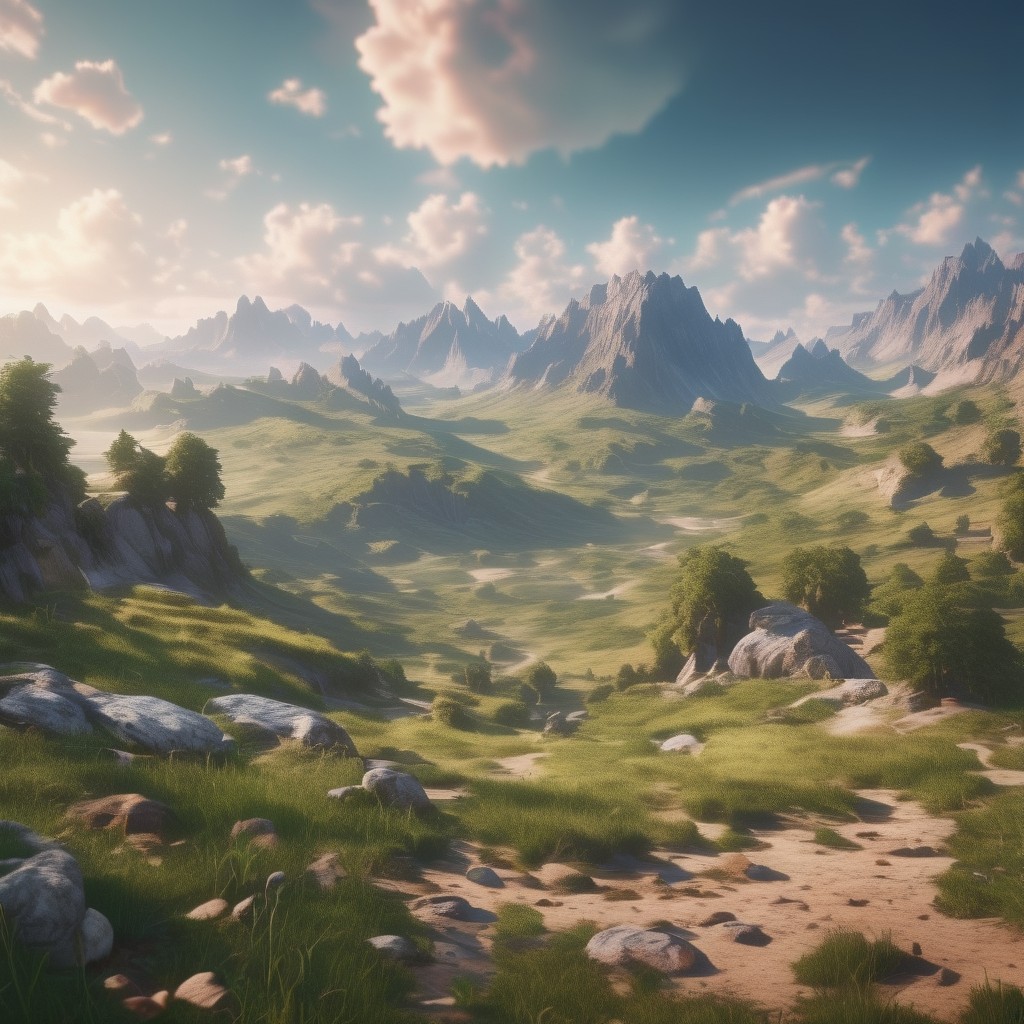} &
    \includegraphics[width=\cw]{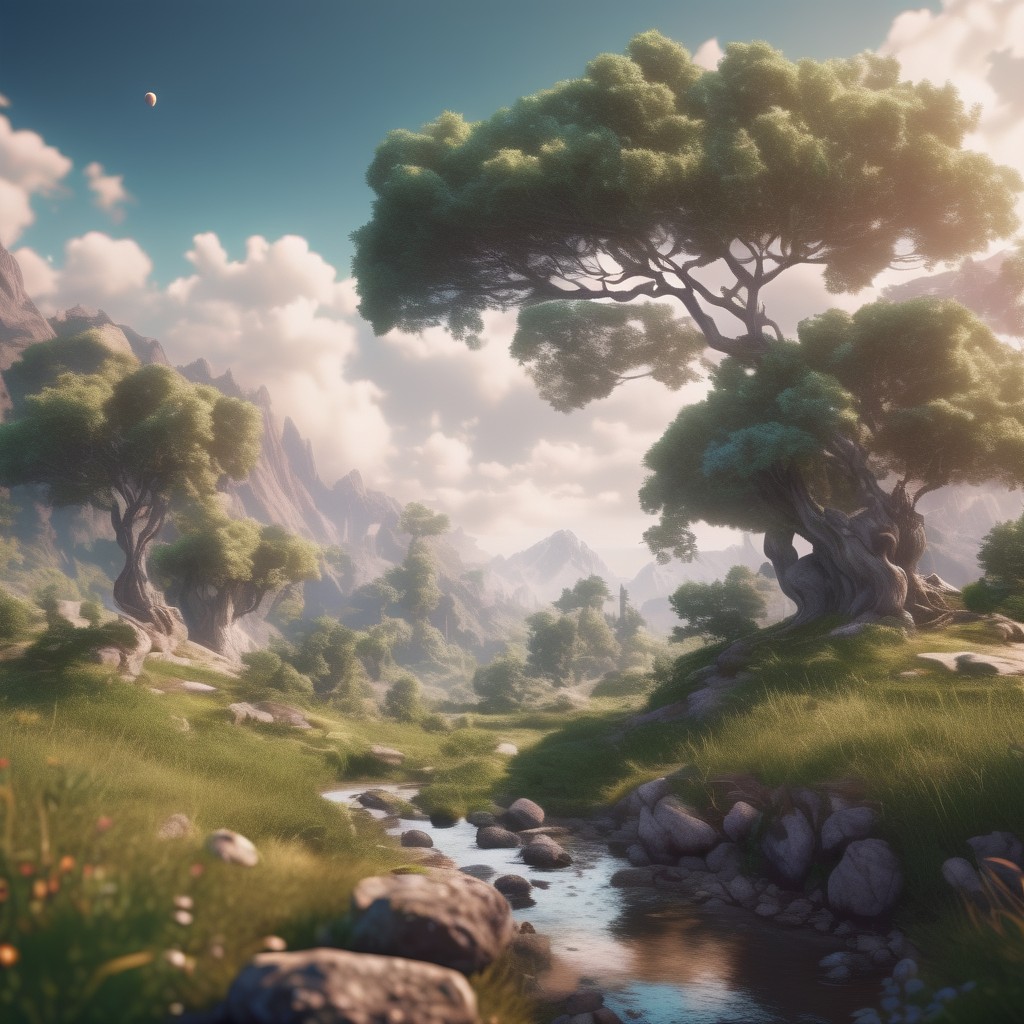} &
    \parbox[c]{0.12\textwidth}{\raggedright\fontsize{7}{8}\selectfont render of dreamy beautiful landscape, dreamy, artger, large scale,..., high detailed, 8 k} &
    \includegraphics[width=\cw]{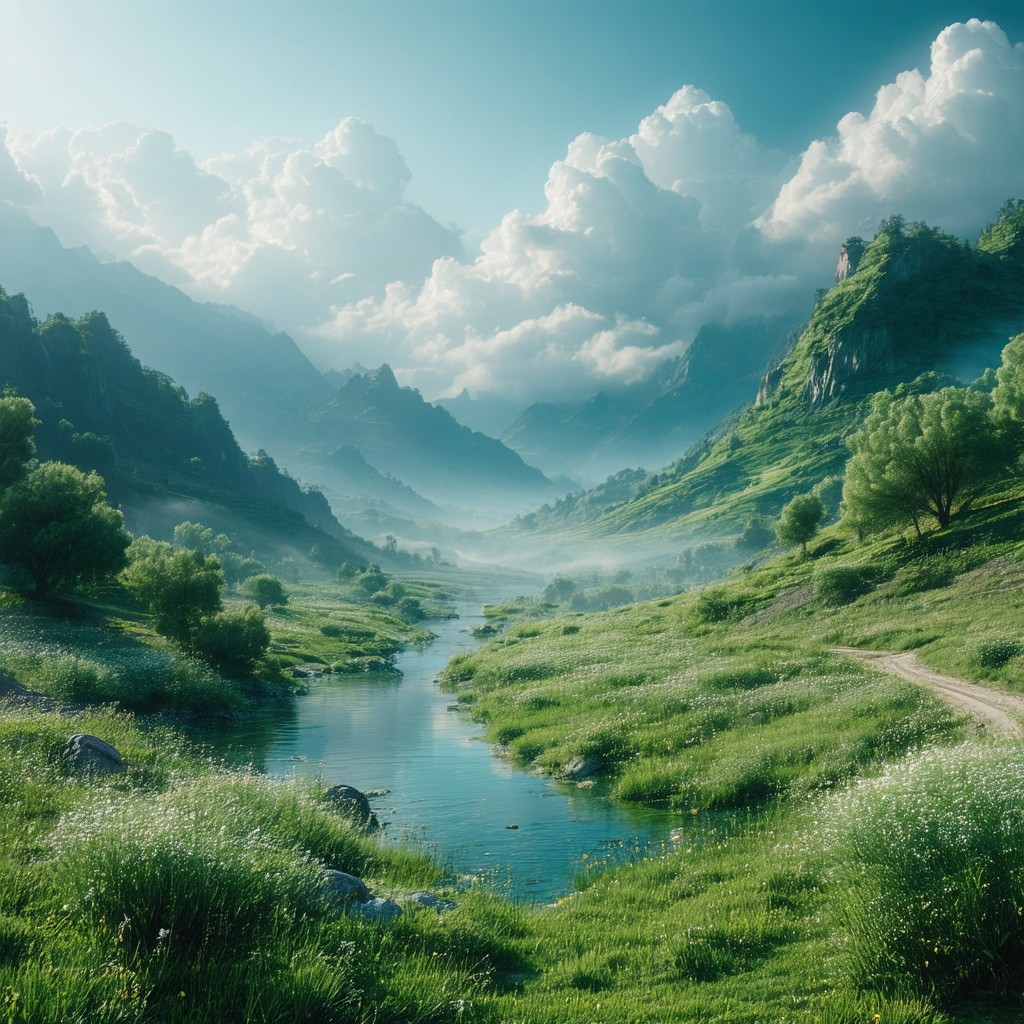} &
    \includegraphics[width=\cw]{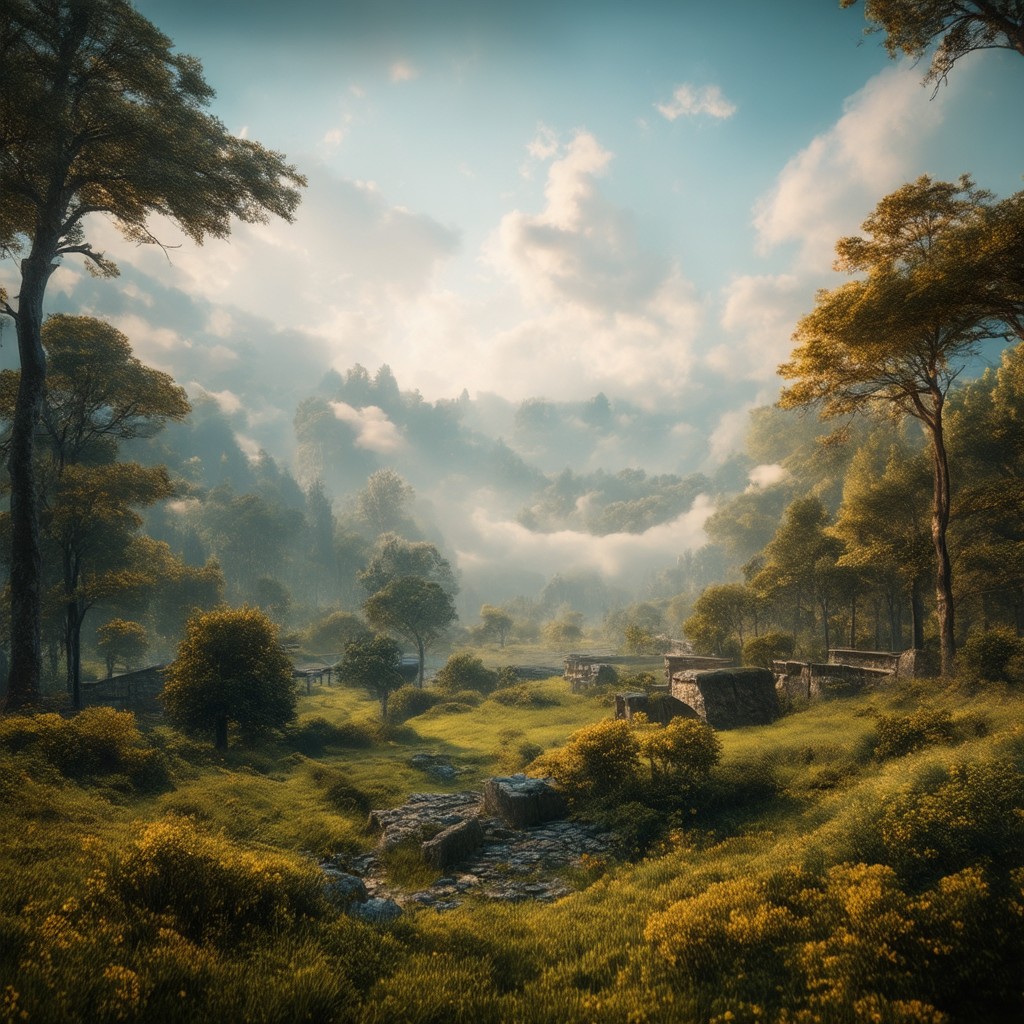} &
    \includegraphics[width=\cw]{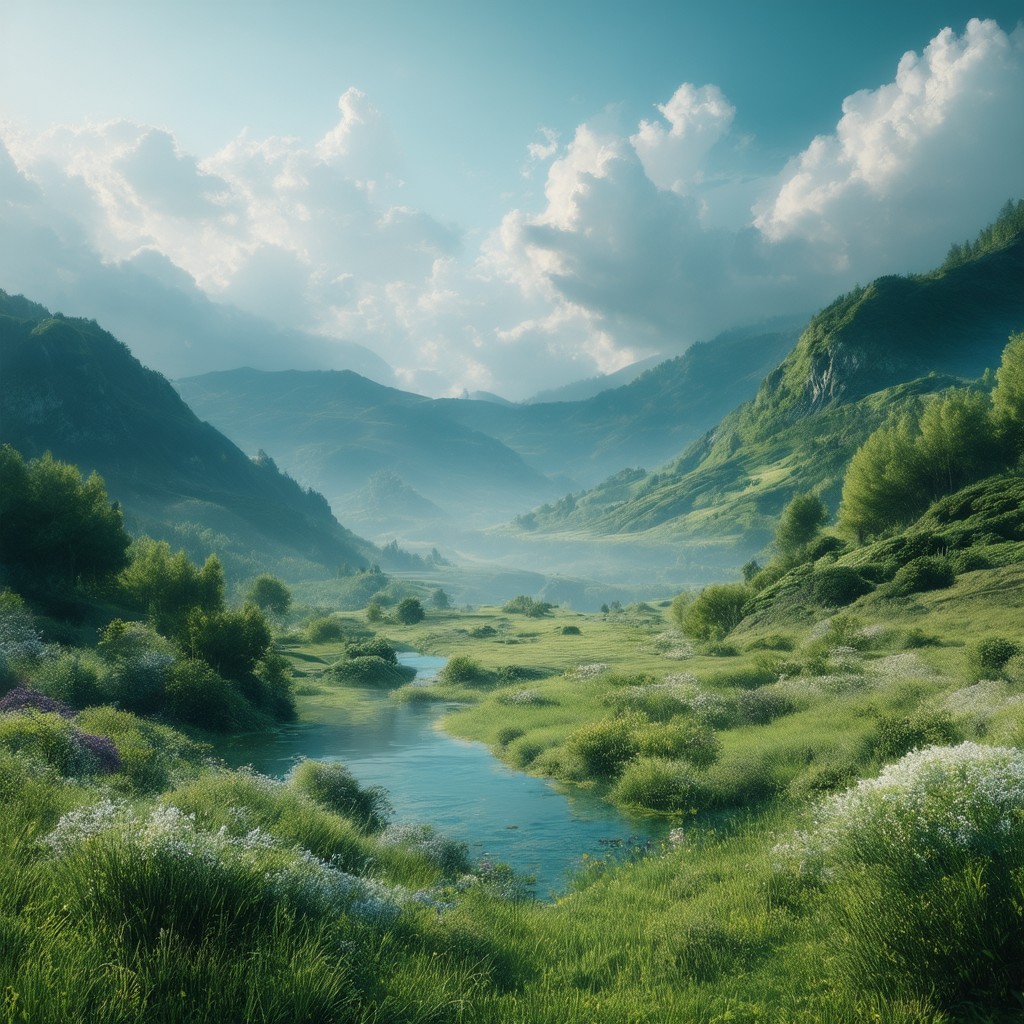} \\[1pt]
    % --- Row 3 ---
    \includegraphics[width=\cw]{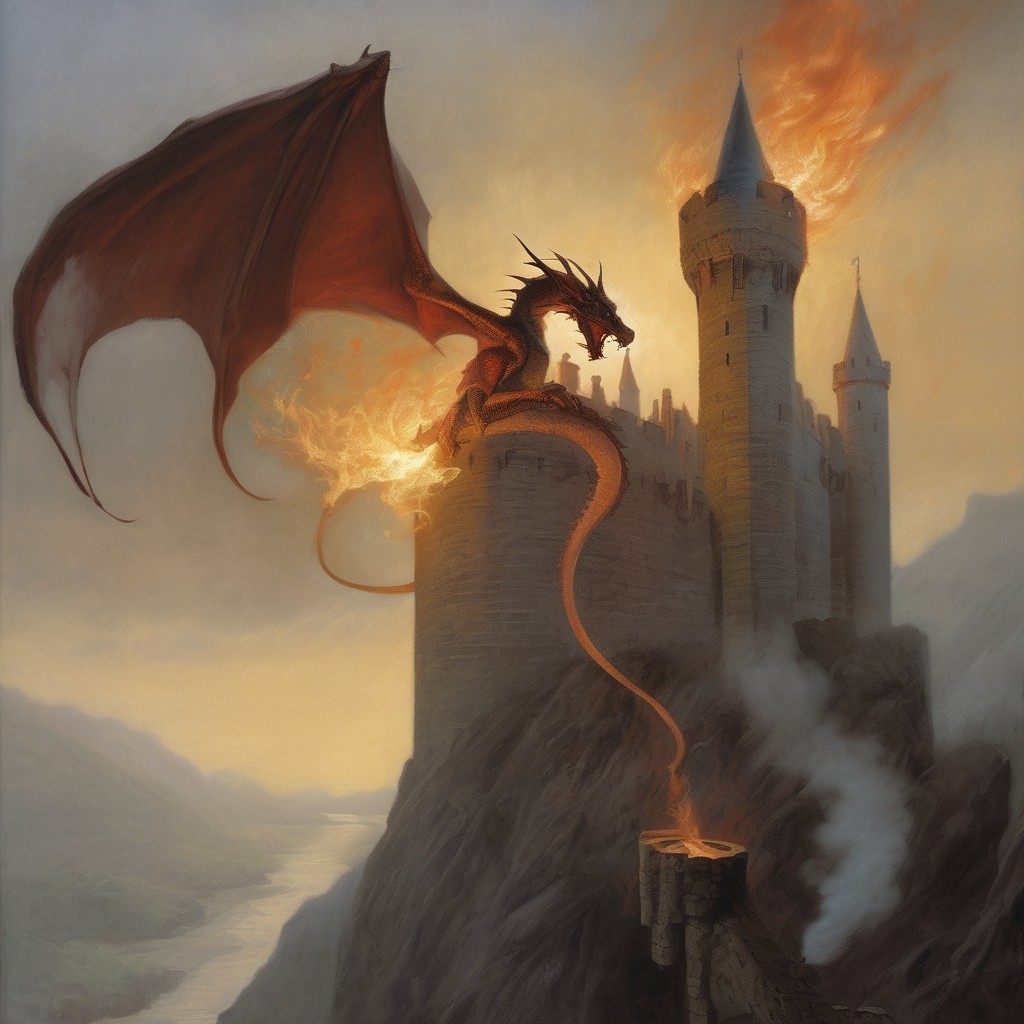} &
    \includegraphics[width=\cw]{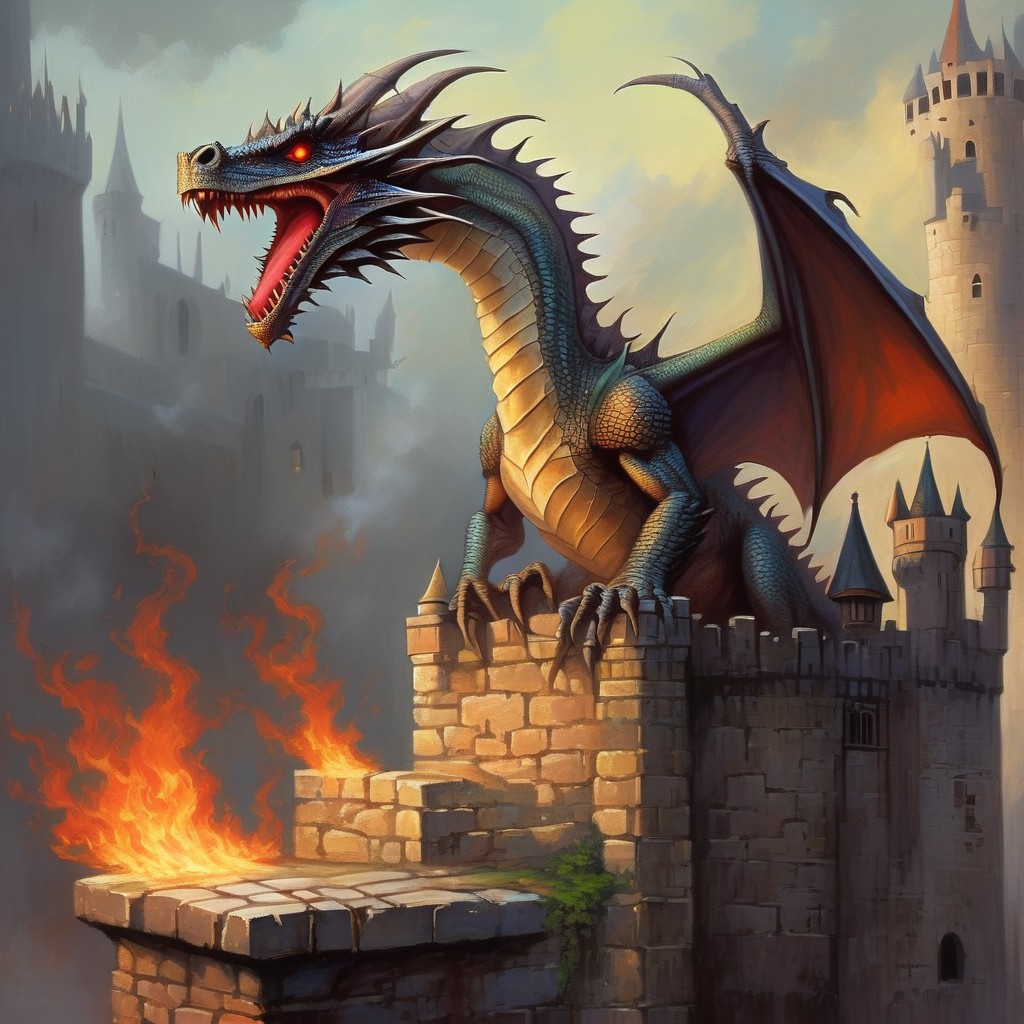} &
    \includegraphics[width=\cw]{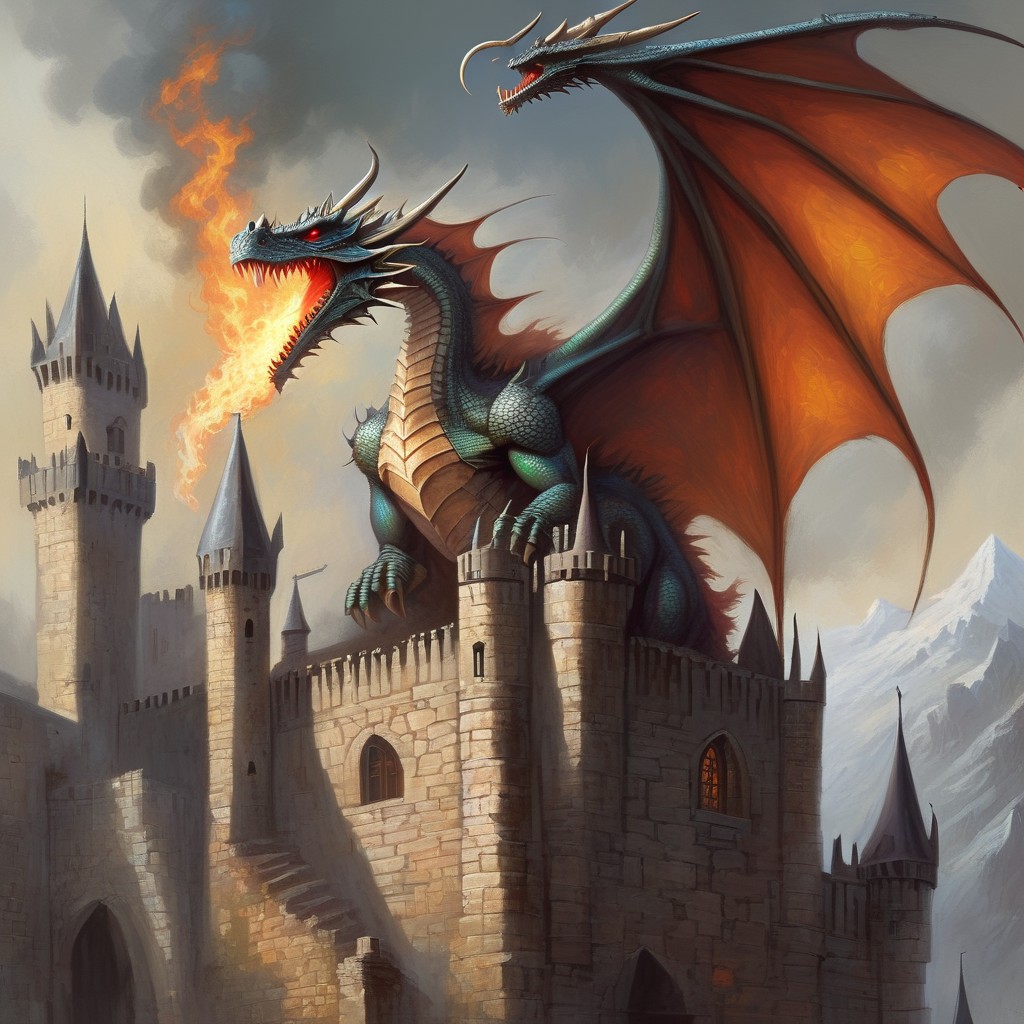} &
    \parbox[c]{0.12\textwidth}{\raggedright\fontsize{7}{8}\selectfont book cover illustration, art by gerald brom, of a flying medieval fantasy dragon...} &
    \includegraphics[width=\cw]{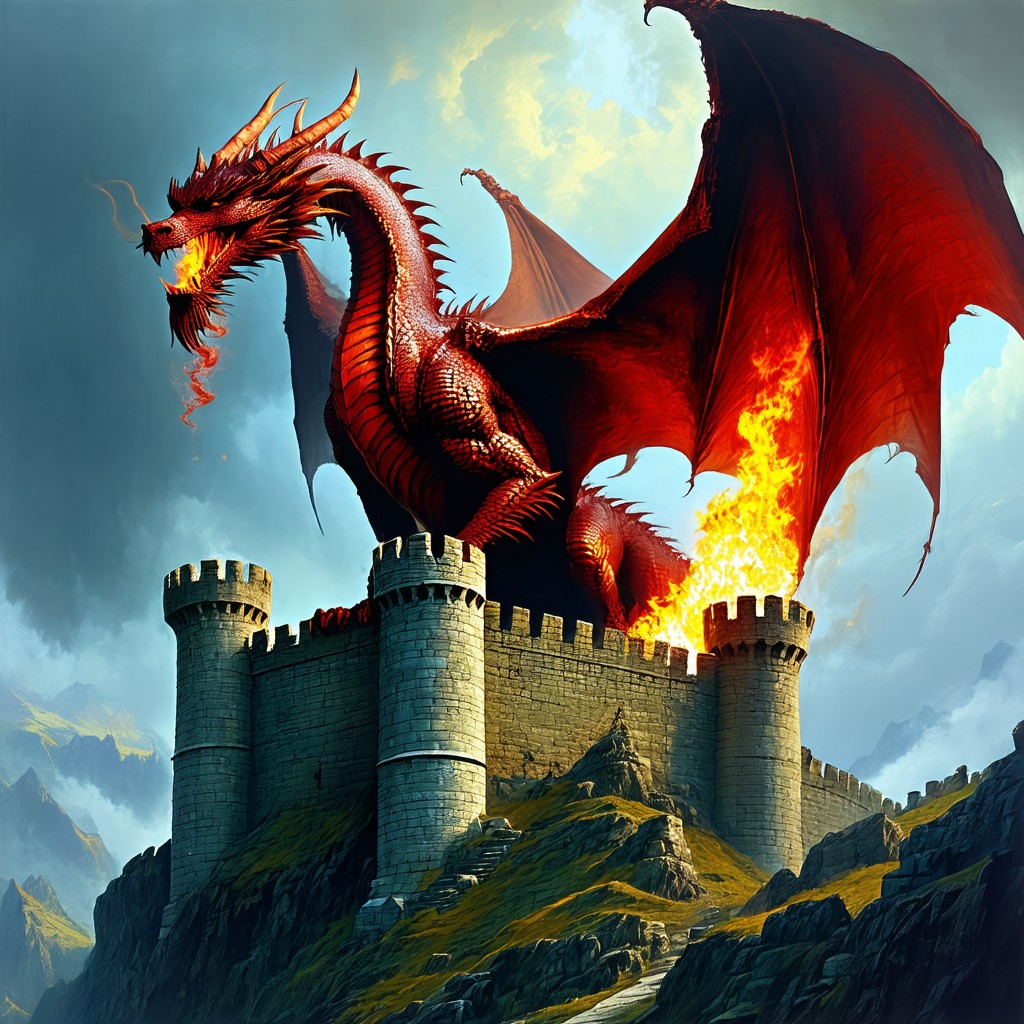} &
    \includegraphics[width=\cw]{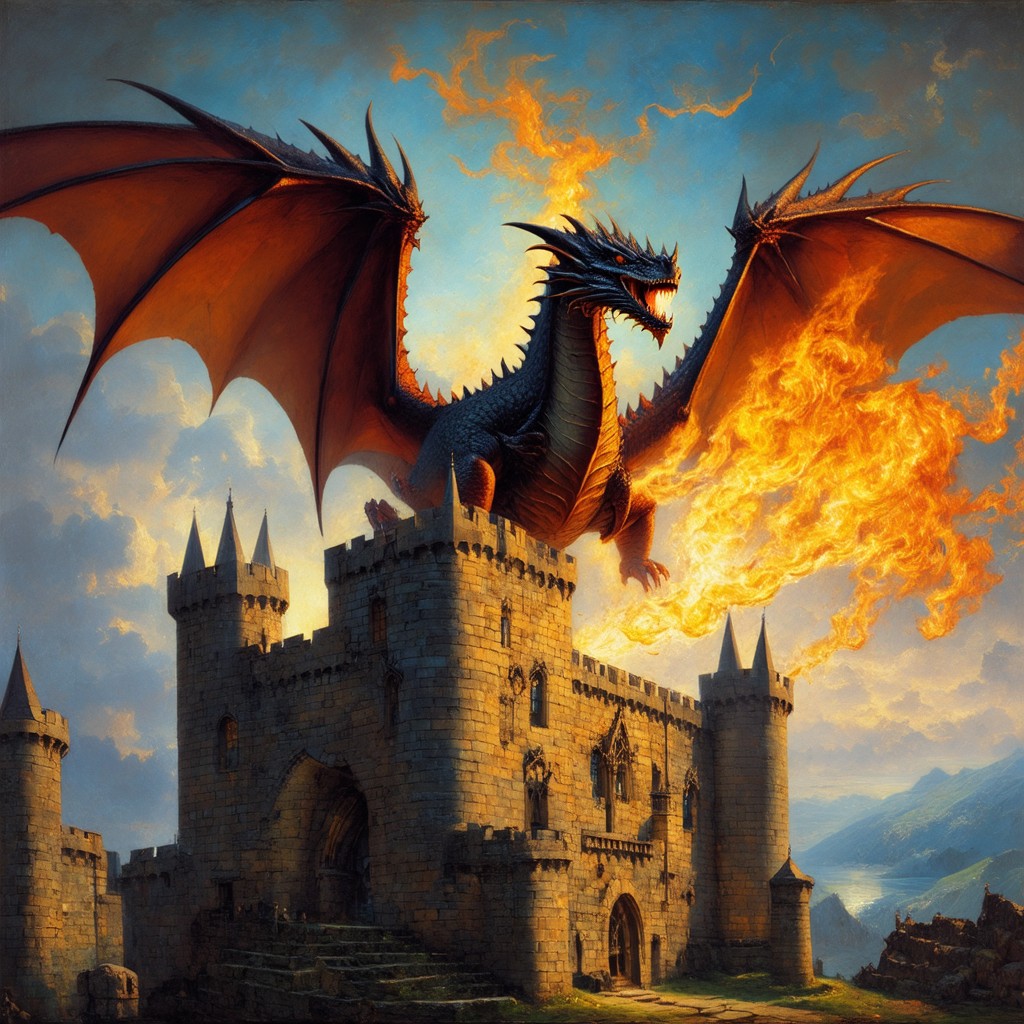} &
    \includegraphics[width=\cw]{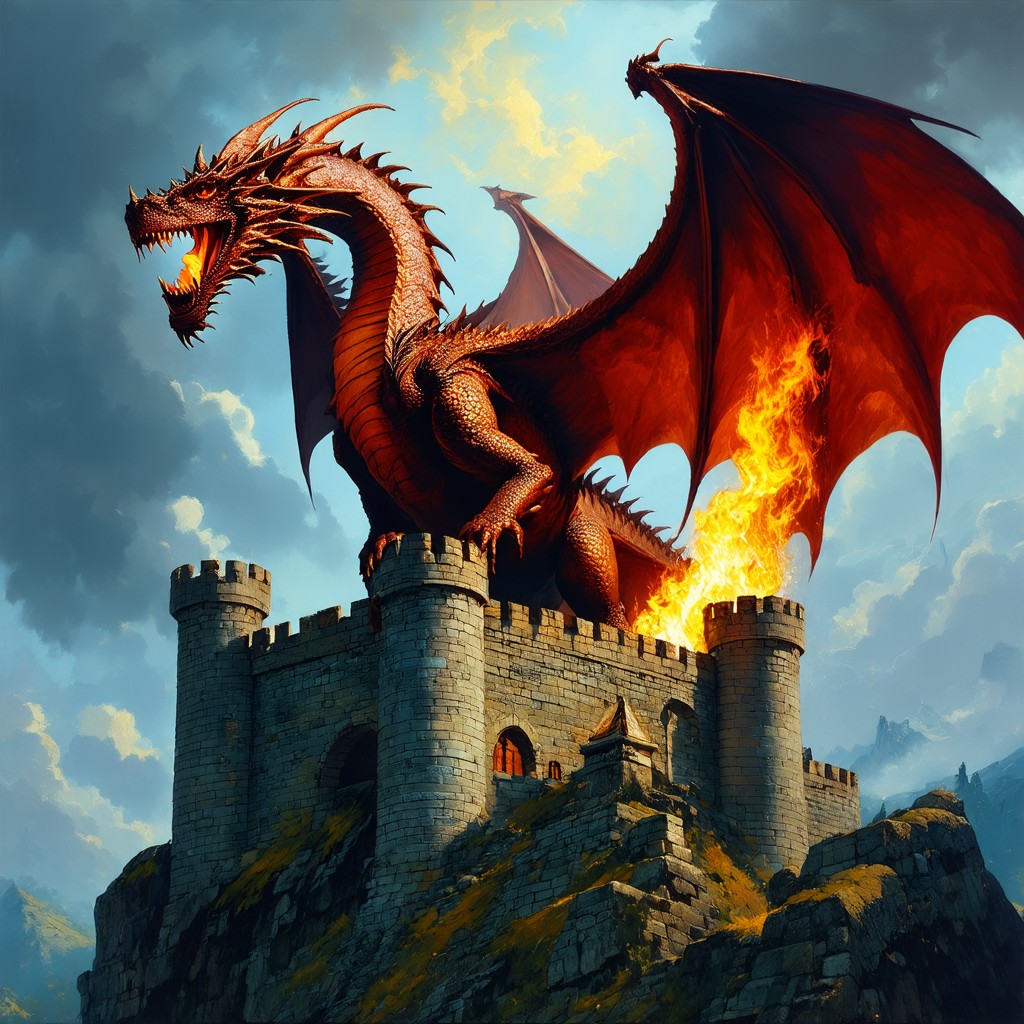} \\[1pt]
    % --- Row 4 ---
    \includegraphics[width=\cw]{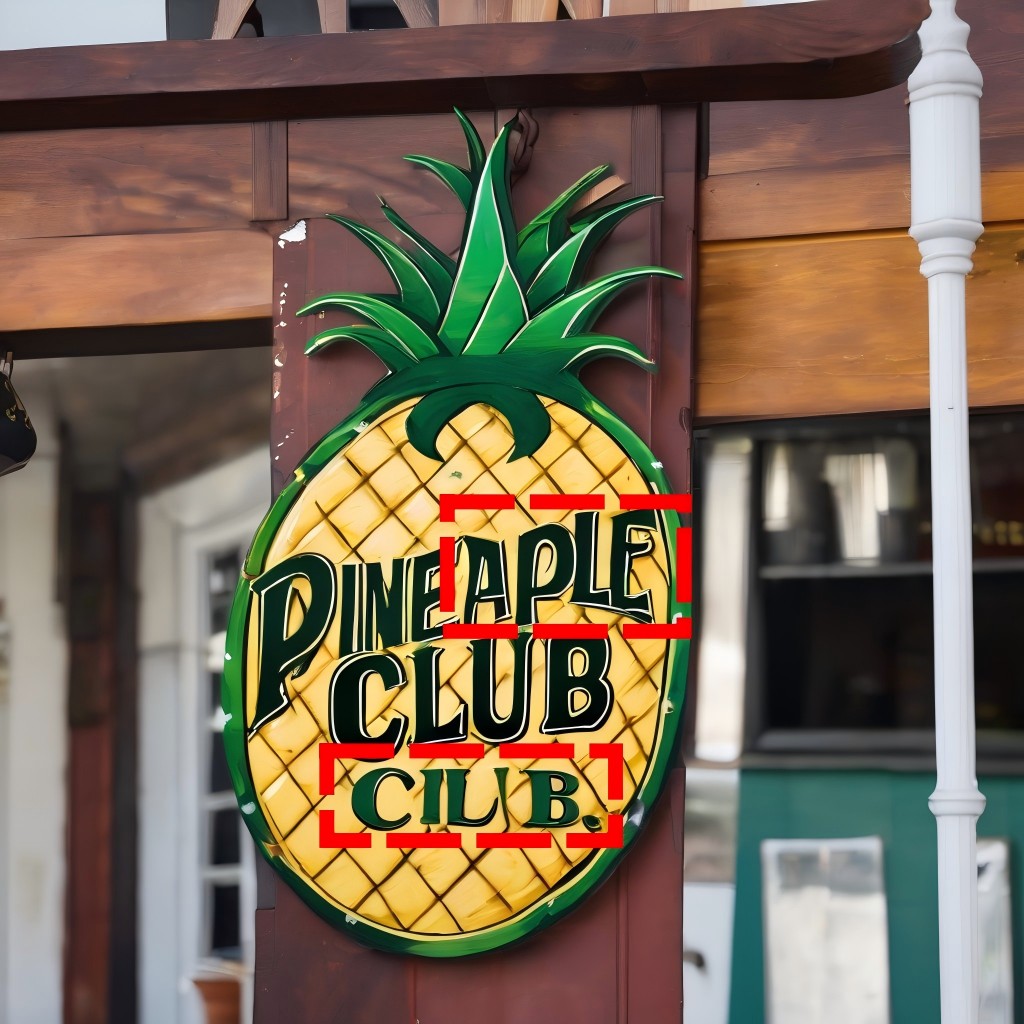} &
    \includegraphics[width=\cw]{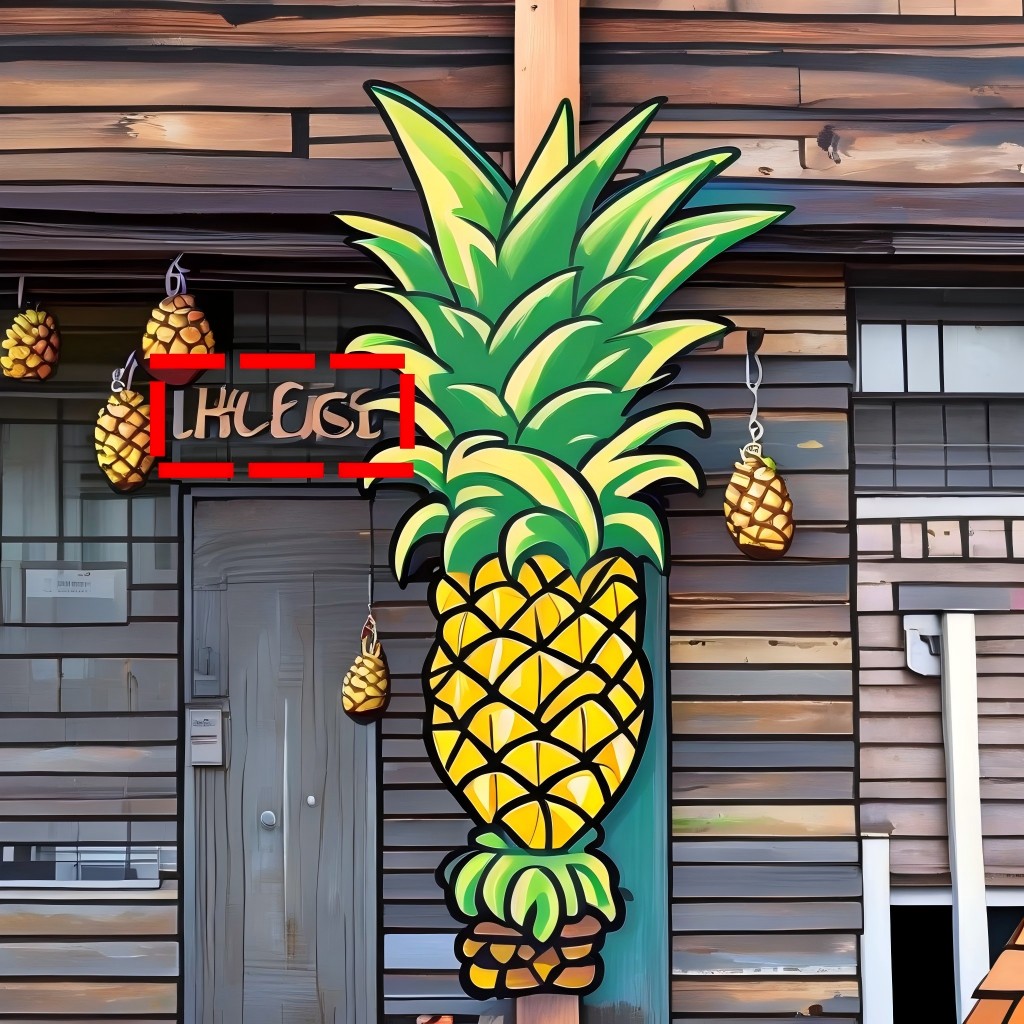} &
    \includegraphics[width=\cw]{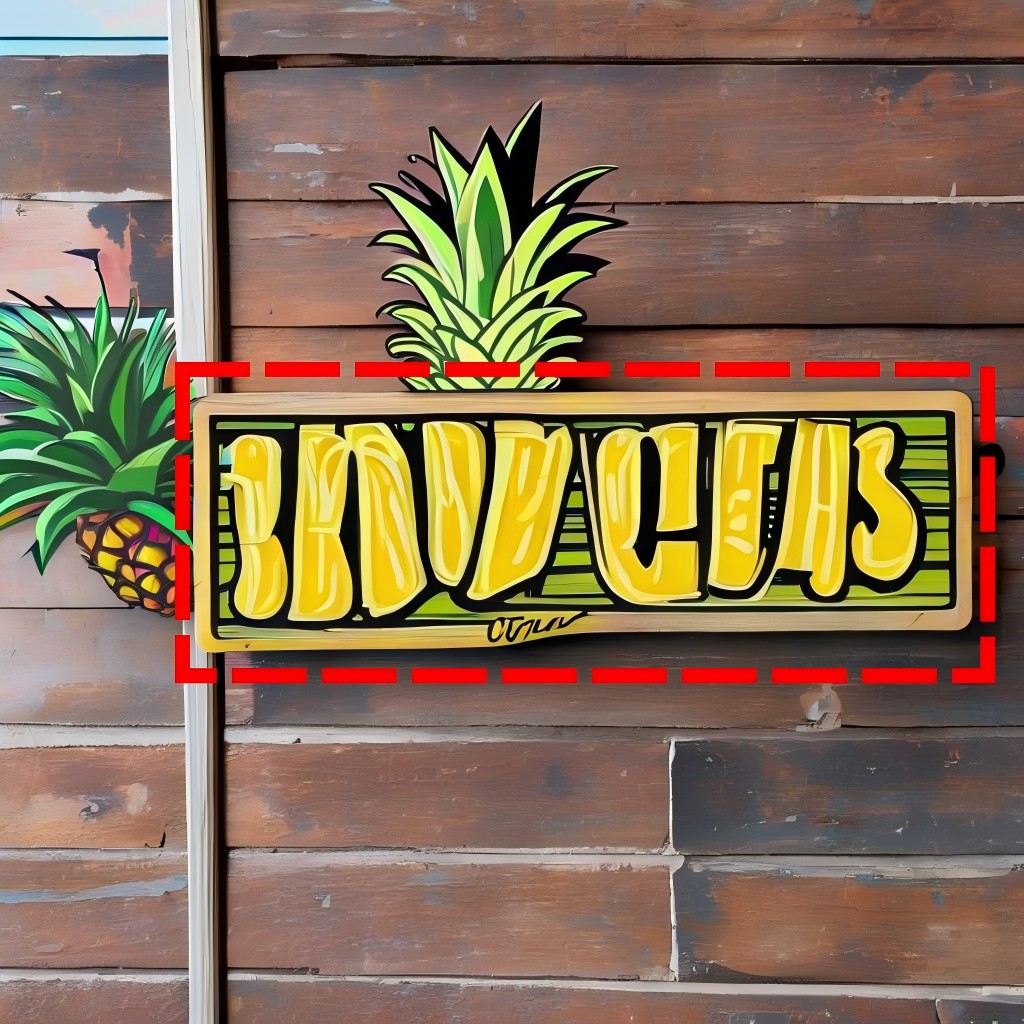} &
    \parbox[c]{0.12\textwidth}{\raggedright\fontsize{7}{8}\selectfont A hand painted wooden ``Pineapple Club'' sign in the shape of a pineapple, hanging outside a bar.} &
    \includegraphics[width=\cw]{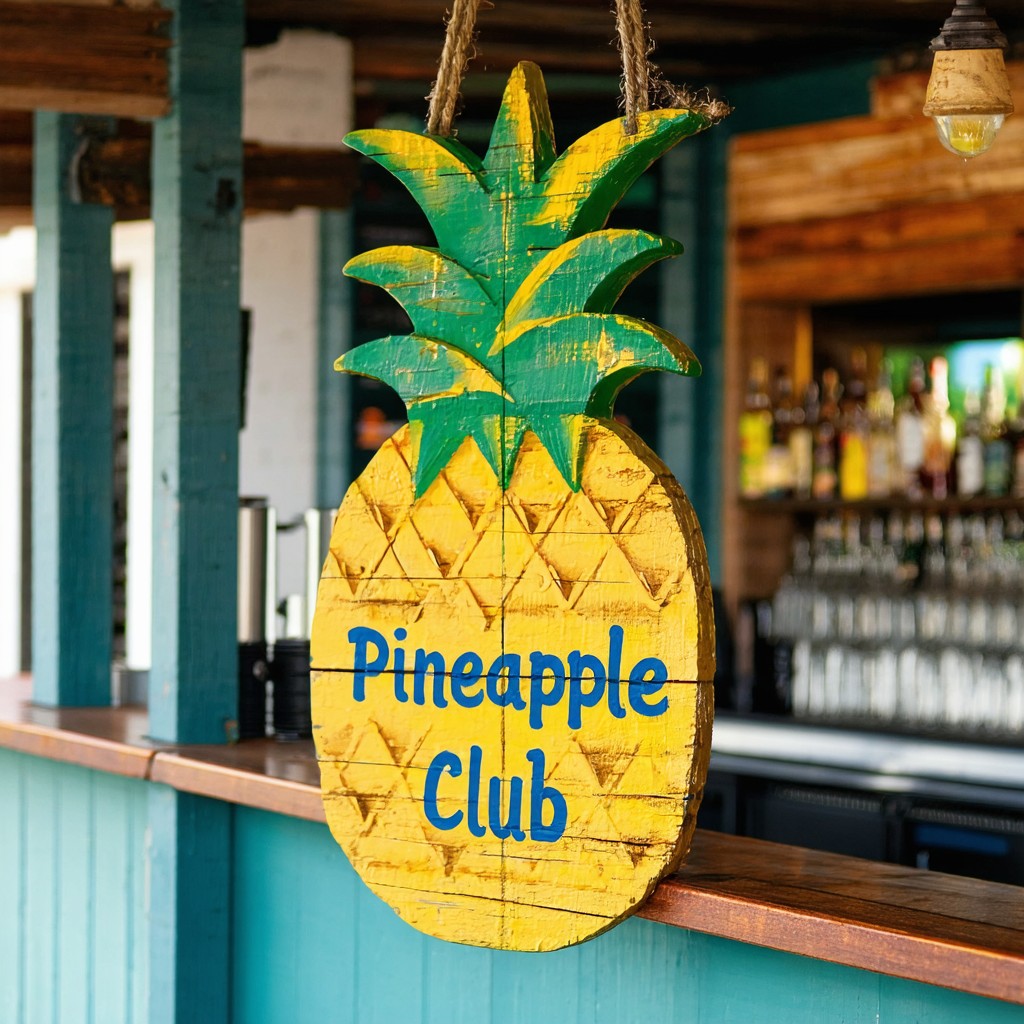} &
    \includegraphics[width=\cw]{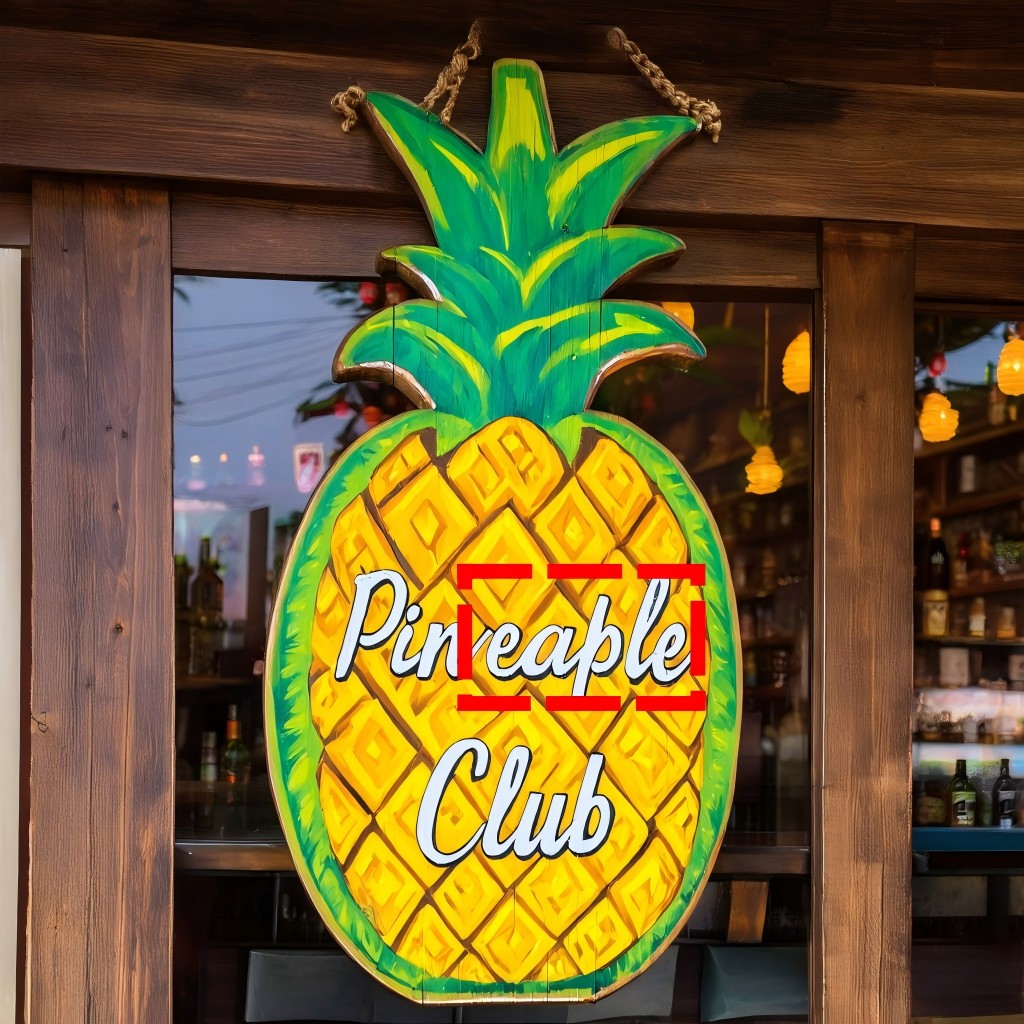} &
    \includegraphics[width=\cw]{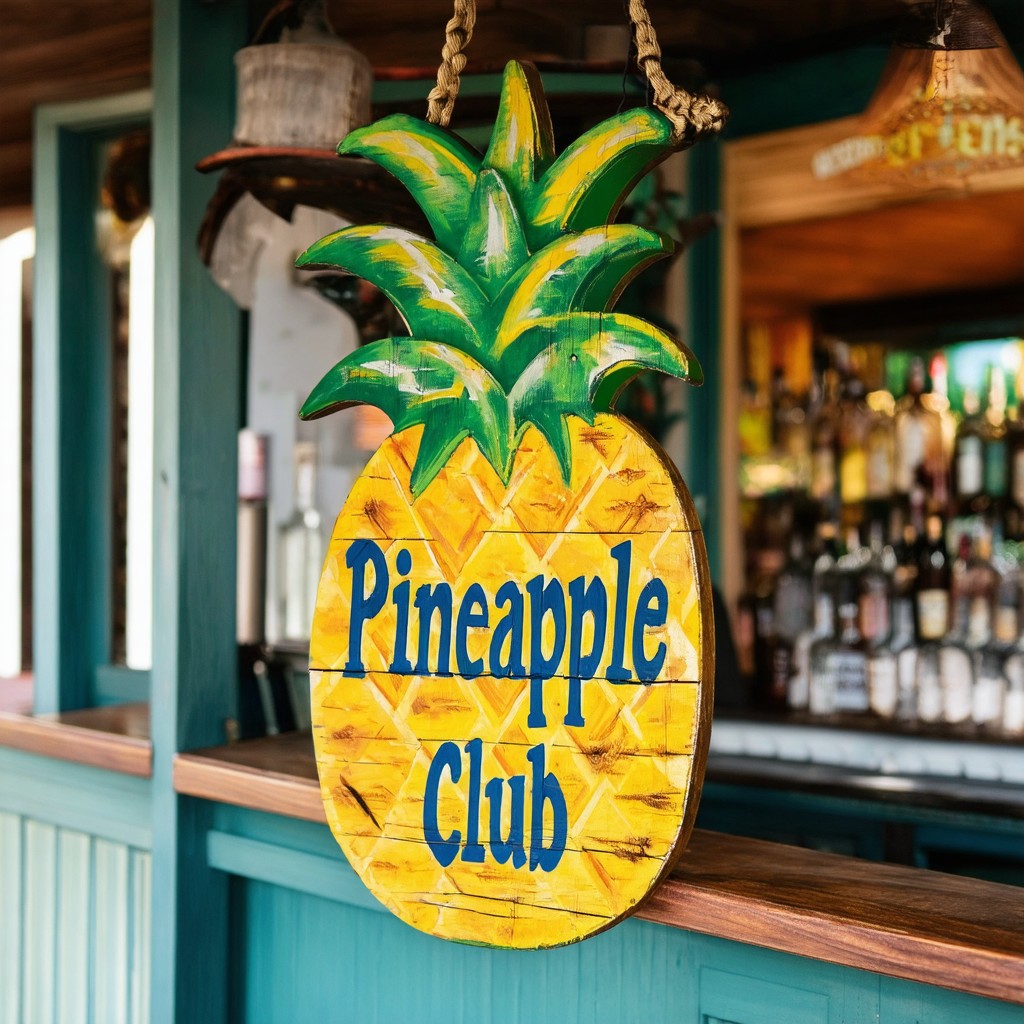} \\[1pt]
    % --- Row 5 ---
    \includegraphics[width=\cw]{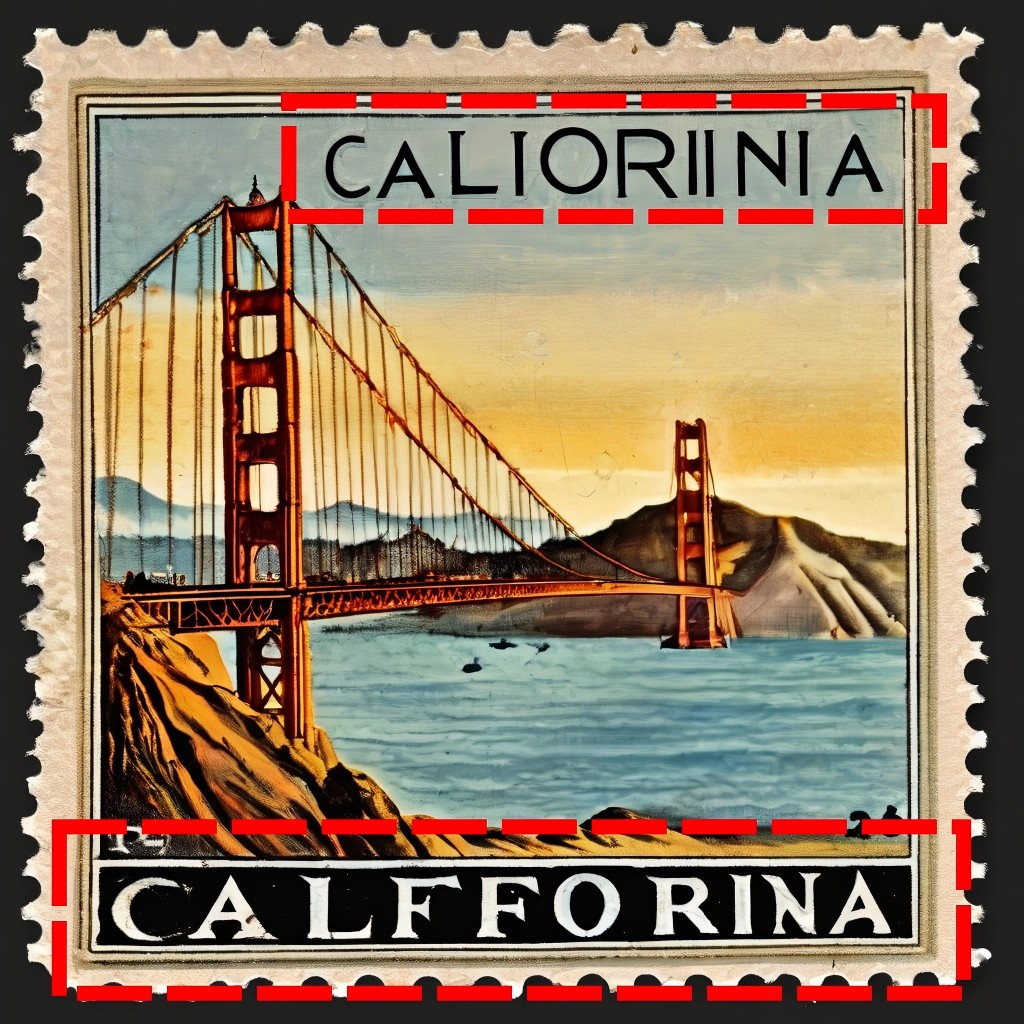} &
    \includegraphics[width=\cw]{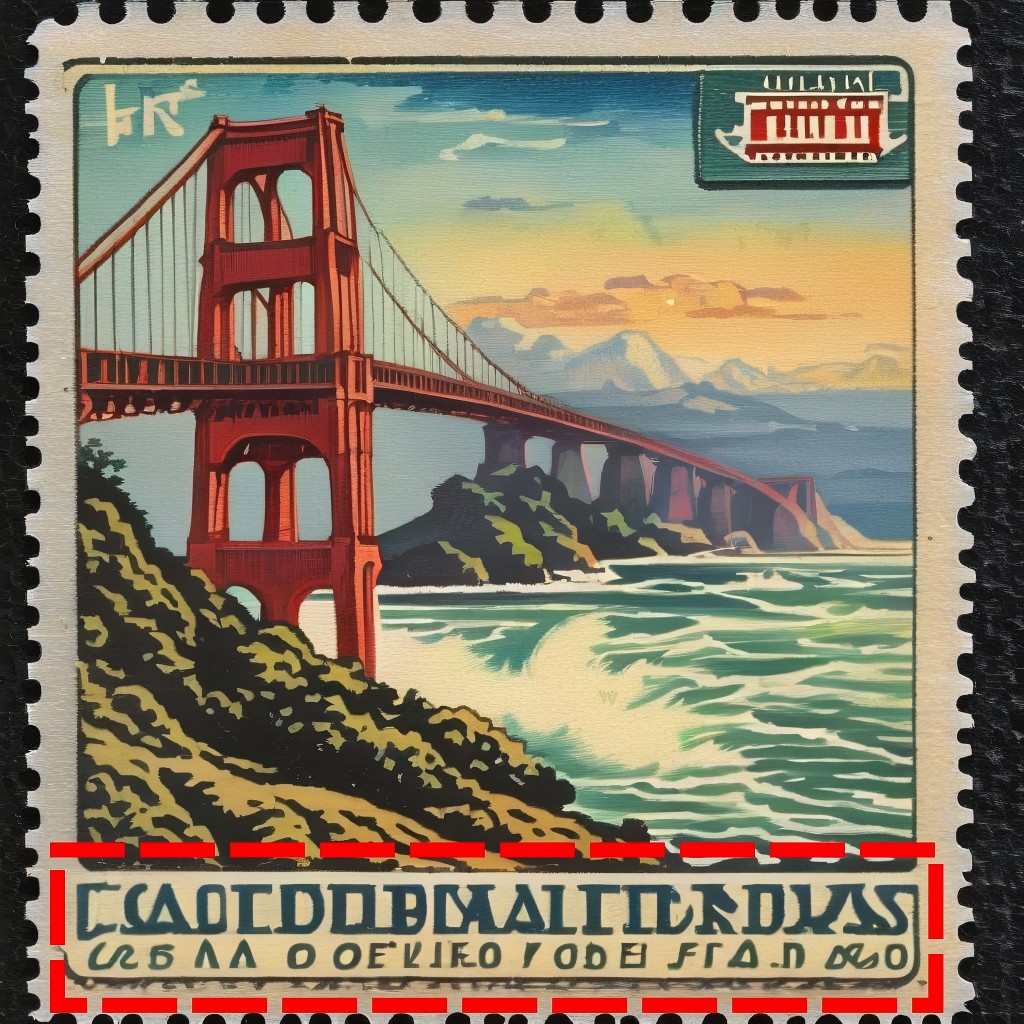} &
    \includegraphics[width=\cw]{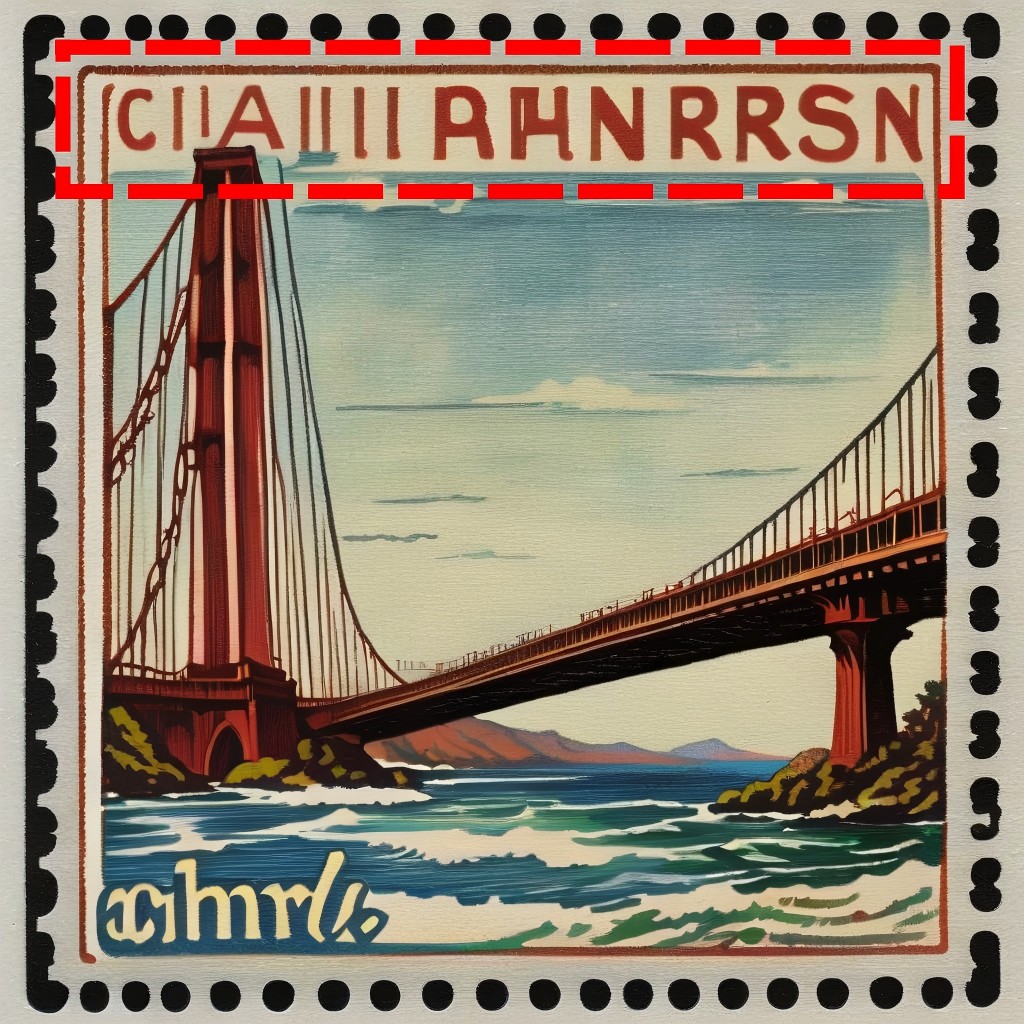} &
    \parbox[c]{0.12\textwidth}{\raggedright\fontsize{7}{8}\selectfont A vintage postage stamp showing a painting of the Golden Gate Bridge and the text ``California''.} &
    \includegraphics[width=\cw]{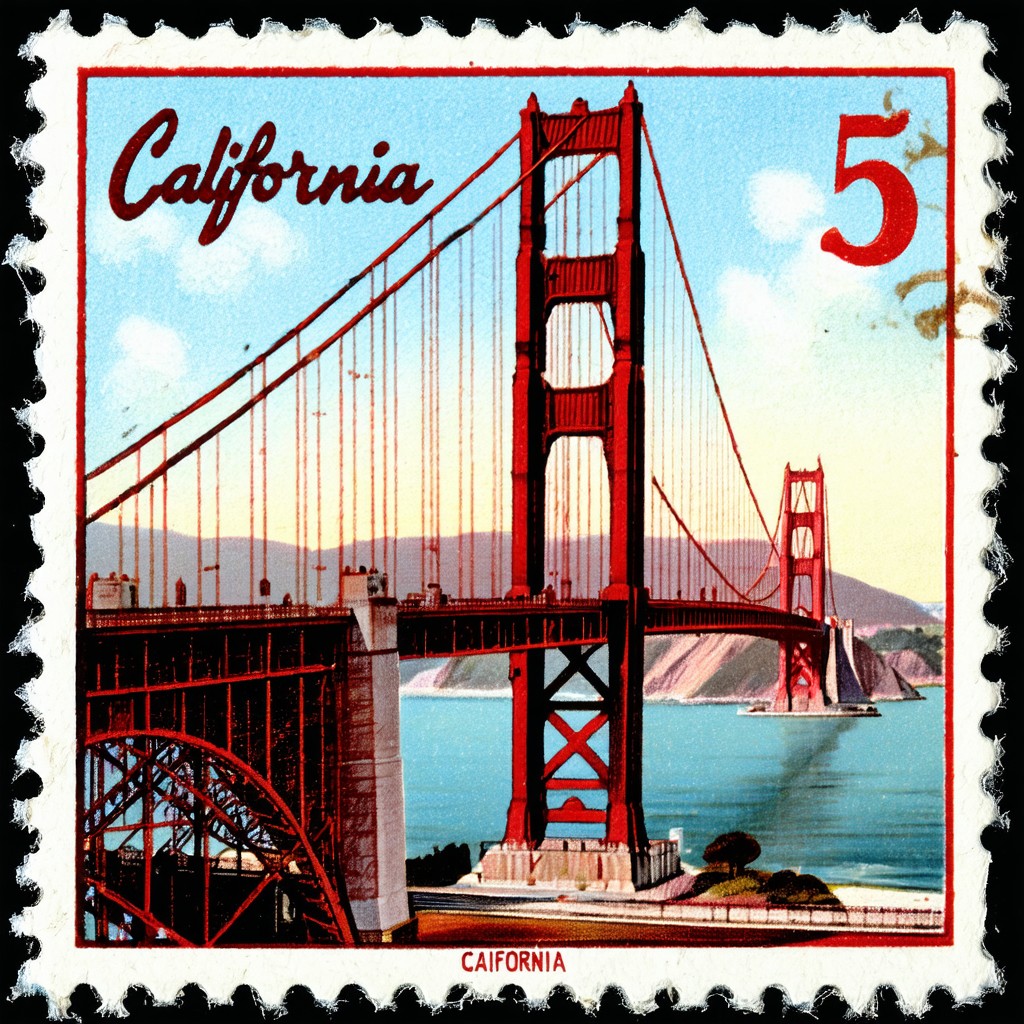} &
    \includegraphics[width=\cw]{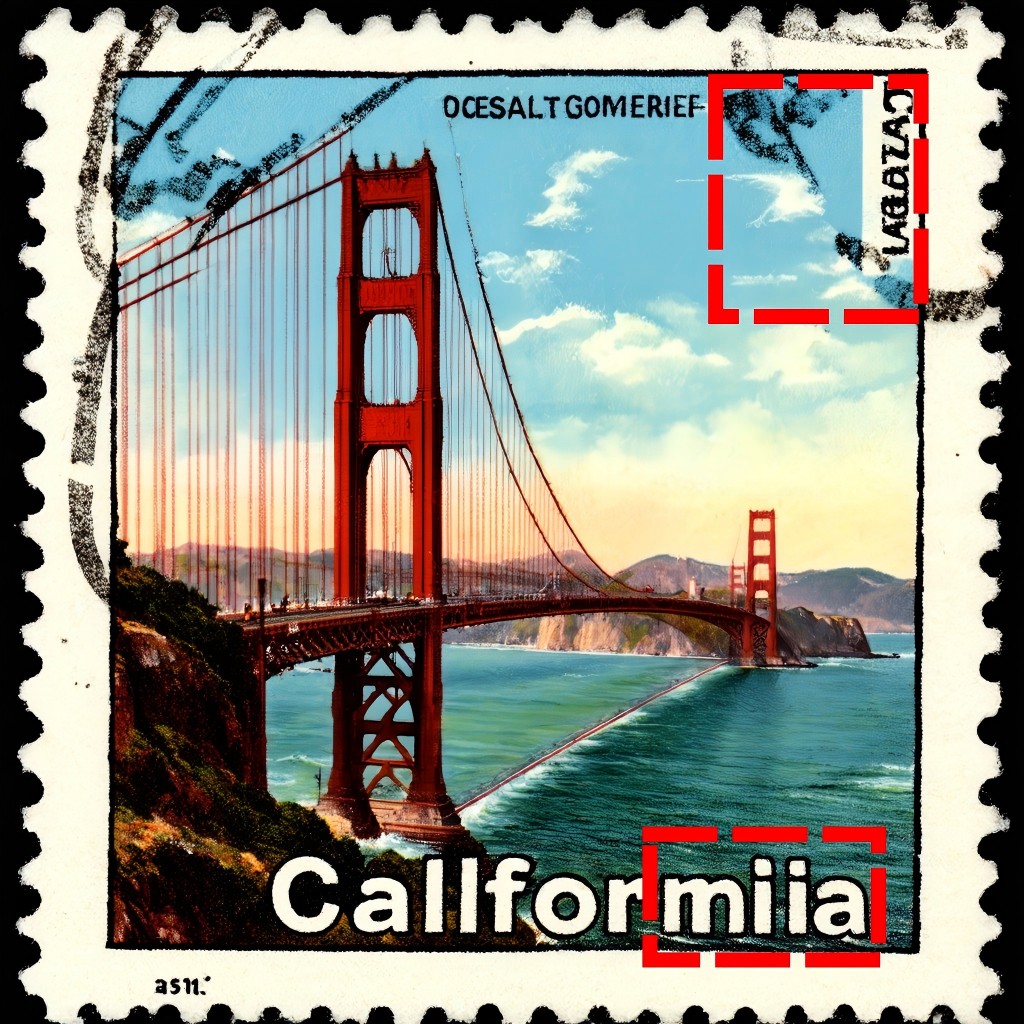} &
    \includegraphics[width=\cw]{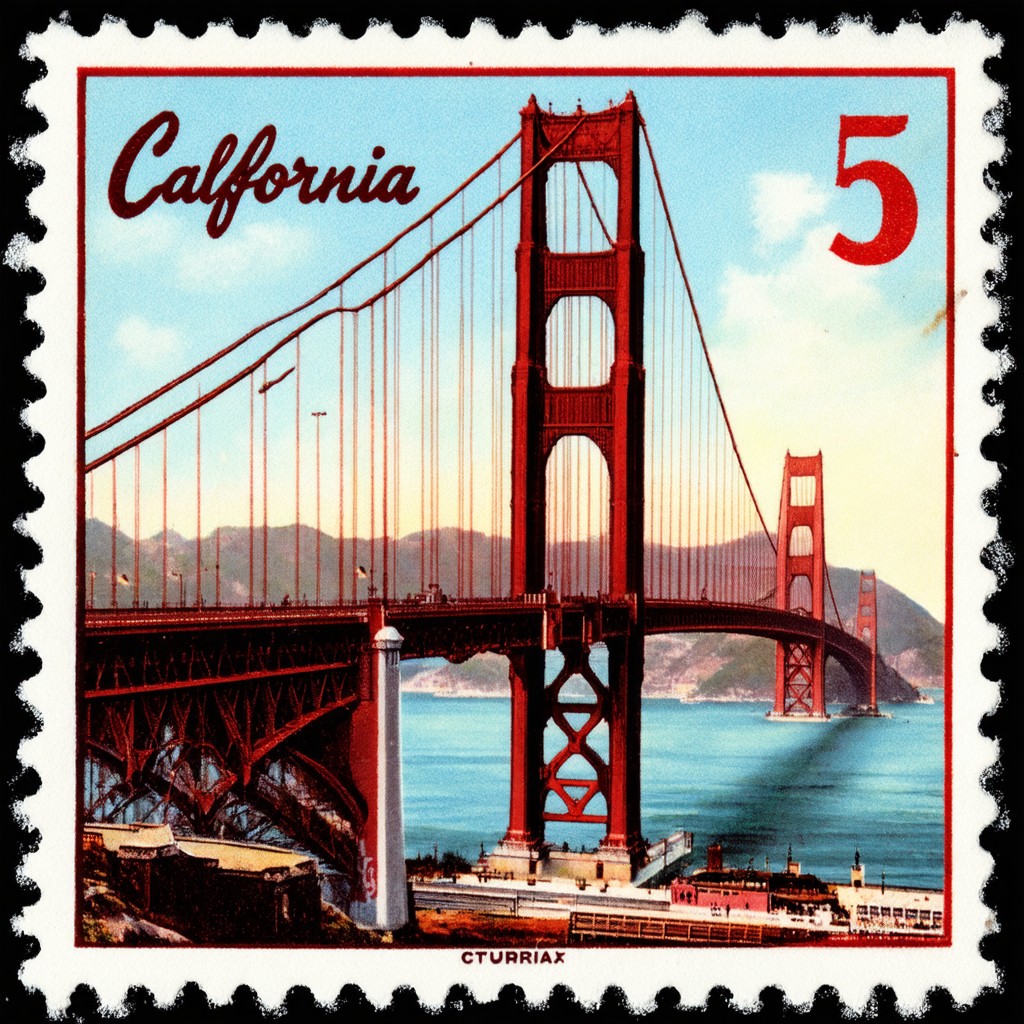} \\
    \end{tabular}
    \caption{Visual comparison across relay configurations.}
    \label{fig:visual_comparison}
\end{figure*}

\textbf{Finding 2: Task type determines the optimal model family, highlighting the necessity of context-aware scheduling.} 
On text-rendering tasks, the SD3 family largely preserves OCR accuracy through the relay process. Conversely, the SDXL family inherently lacks strong text-rendering capabilities, and we observe that this architectural limitation is further amplified during relay inference, leading to degraded text readability. Moreover, this task-dependent variance precisely underscores the necessity of our context-aware online scheduler. By recognizing prompt requirements, the scheduler intelligently routes text-heavy requests to the capable SD3 family while reserving the highly efficient SDXL relay configurations for general visual tasks.

\subsection{Parameter Sensitivity Analysis (RQ1)}

\textbf{Qualitative Verification.} We first visually inspect the generated images. Figure~\ref{fig:visual_comparison} demonstrates that for general visuals, SDXL relay variants produce coherent images at dramatically higher speeds, though all SDXL configurations consistently fail at text rendering. Conversely, SD3 relay variants remain visually close to the full SD3.5 Large model. Notably, they avoid the visible text degradation suffered by the standalone SD3.5 Medium baseline, visually confirming that leveraging a large model's semantic foundation yields strictly better results than using a smaller standalone model.

\begin{figure*}[t]
    \centering
    \newcommand{\hw}{0.32\textwidth}

    \begin{subfigure}{\hw}
        \centering
        \includegraphics[width=\textwidth]{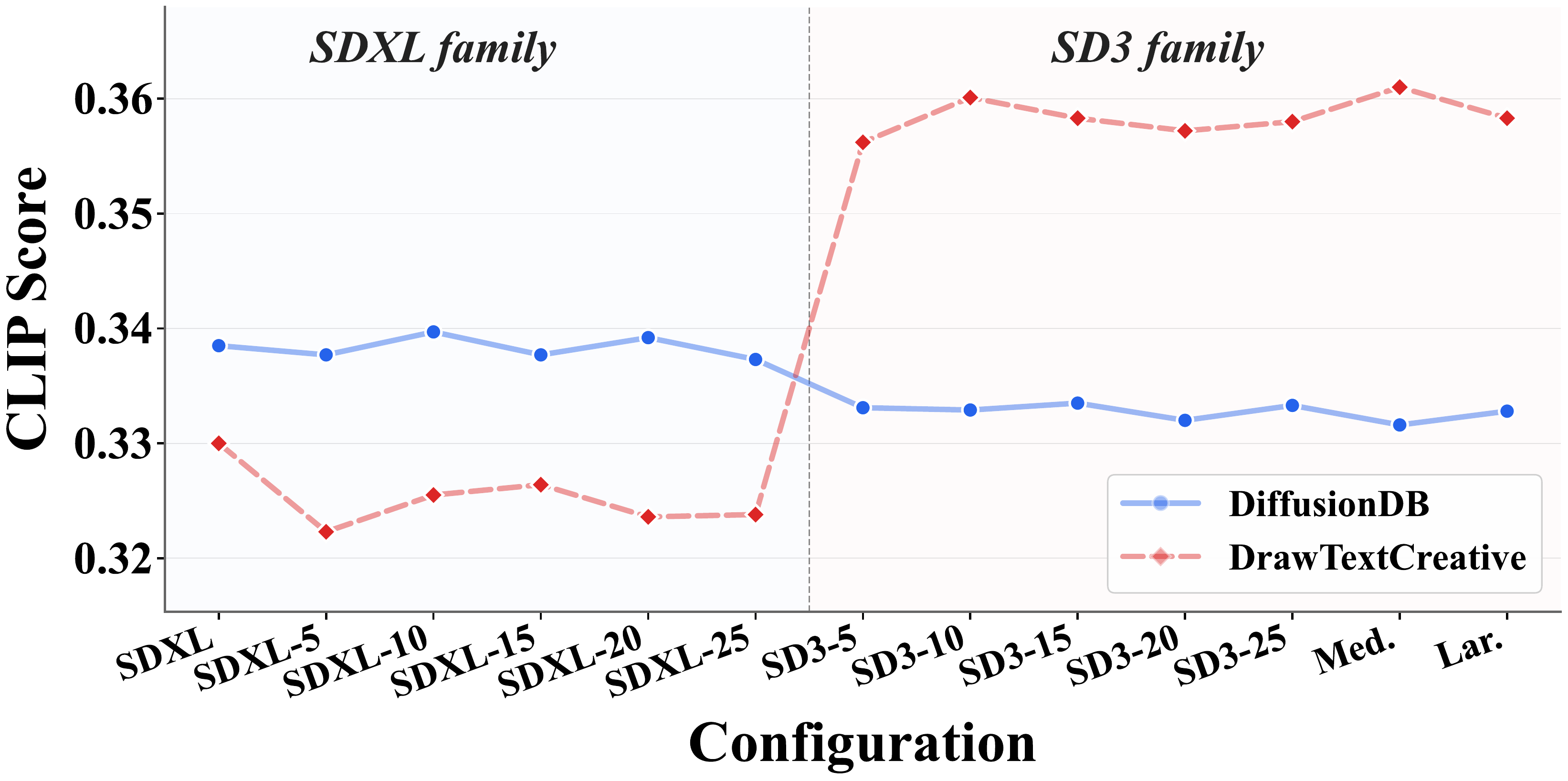}
        \caption{CLIP Score}
        \label{fig:clip_score}
    \end{subfigure}%
    \hfill
    \begin{subfigure}{\hw}
        \centering
        \includegraphics[width=\textwidth]{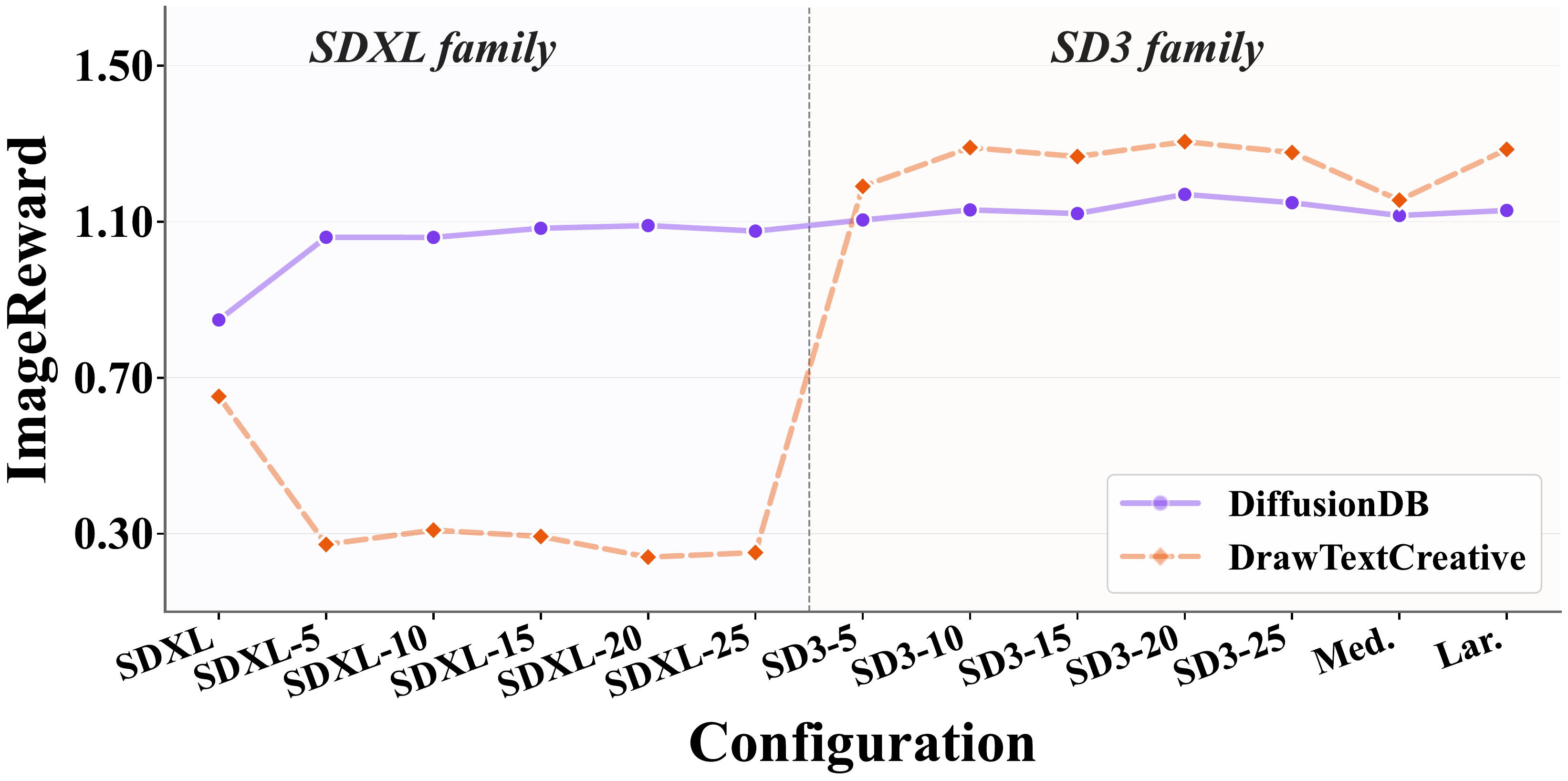}
        \caption{ImageReward}
        \label{fig:image_reward}
    \end{subfigure}%
    \hfill
    \begin{subfigure}{\hw}
        \centering
        \includegraphics[width=\textwidth]{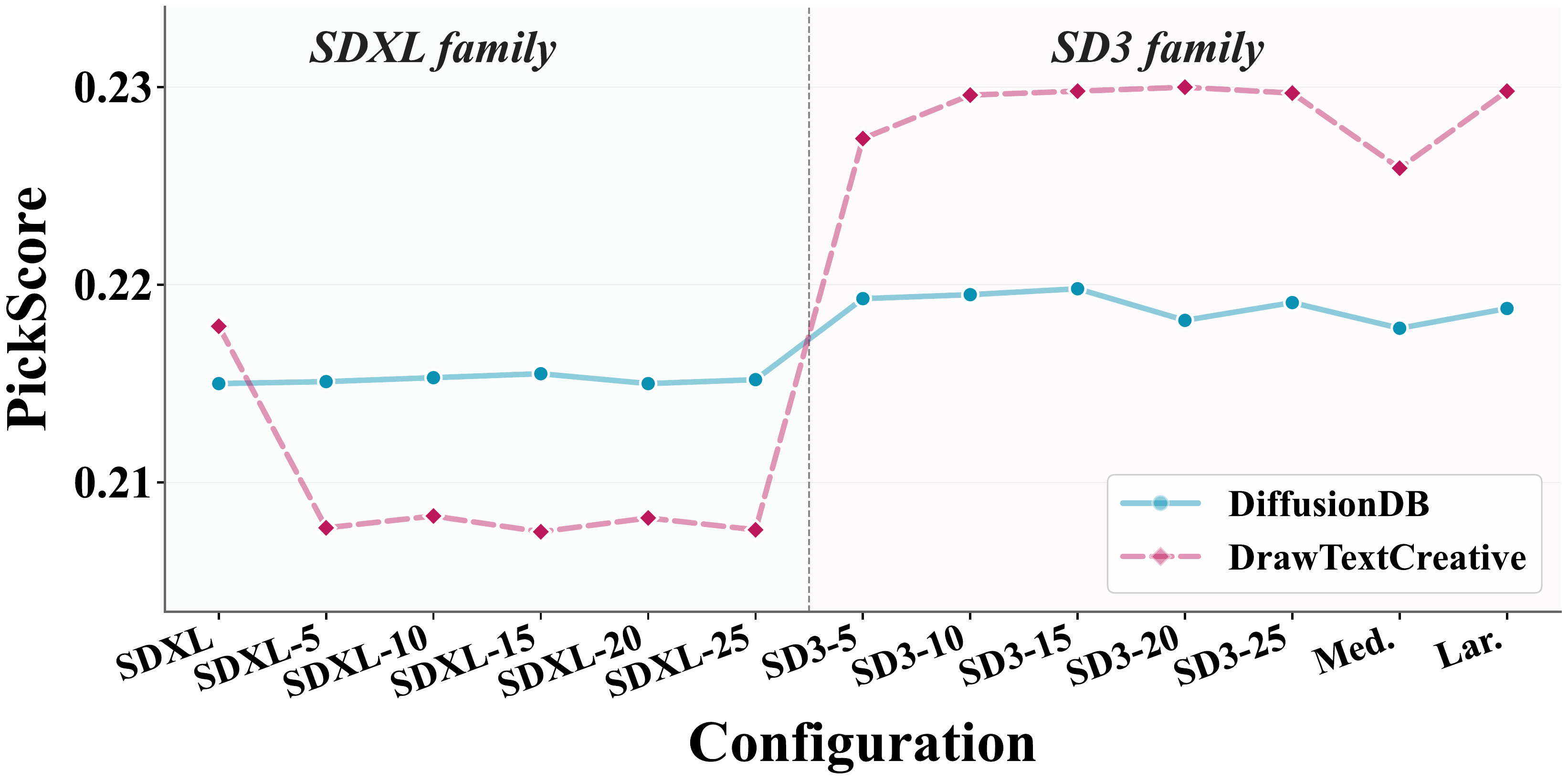}
        \caption{PickScore}
        \label{fig:pick_score}
    \end{subfigure}

    \caption{Quality metrics vs.\ relay step on both datasets. \textit{Family}-$s$ denotes a relay configuration where the large model handles the first $s$ denoising steps before handing off to the small model (e.g., SD3-5); SDXL, SD3.5 Medium, and SD3.5 Large are standalone baselines.}
    \label{fig:metrics_grid}
\end{figure*}

To formalize these observations and characterize the quality--latency tradeoff, we quantitatively analyze the impact of relay step $s$ on generation quality. Figure~\ref{fig:metrics_grid} presents three quality metrics across all ten relay configurations and the three full-model baselines (SDXL, SD3.5 Medium, and SD3.5 Large).

The most prominent pattern is the inter-family gap on text-rendering tasks. On DrawTextCreative, all metrics exhibit a sharp jump at the SDXL-25/SD3-5 boundary. This corroborates our visual findings, confirming that the SDXL-Vega family fundamentally lacks text-rendering capability regardless of relay step, whereas SD3 relay preserves it. On DiffusionDB, both families perform comparably.

Within each family, the relay step has a relatively modest effect. These differential sensitivities justify RISE's two-level scheduling: task type dominates family selection, while the relay step fine-tunes quality within a family.

\subsection{Scheduling Policy Evaluation (RQ2)}

We evaluate the RISE scheduler against four baseline strategies under a multi-tenant service workload. Tasks are drawn from both datasets via stratified sampling into three experimental groups, with task arrivals following a Poisson process with average time $\mu{=}9$\,s. Each scheduling policy processes the same task sequence and selects from 11 configurations across the 4 device pools. All learned schedulers are trained offline on the same training set and evaluated on the test set.

\begin{figure*}[t]
    \centering
    \begin{subfigure}{0.225\textwidth}
        \centering
        \includegraphics[width=\textwidth]{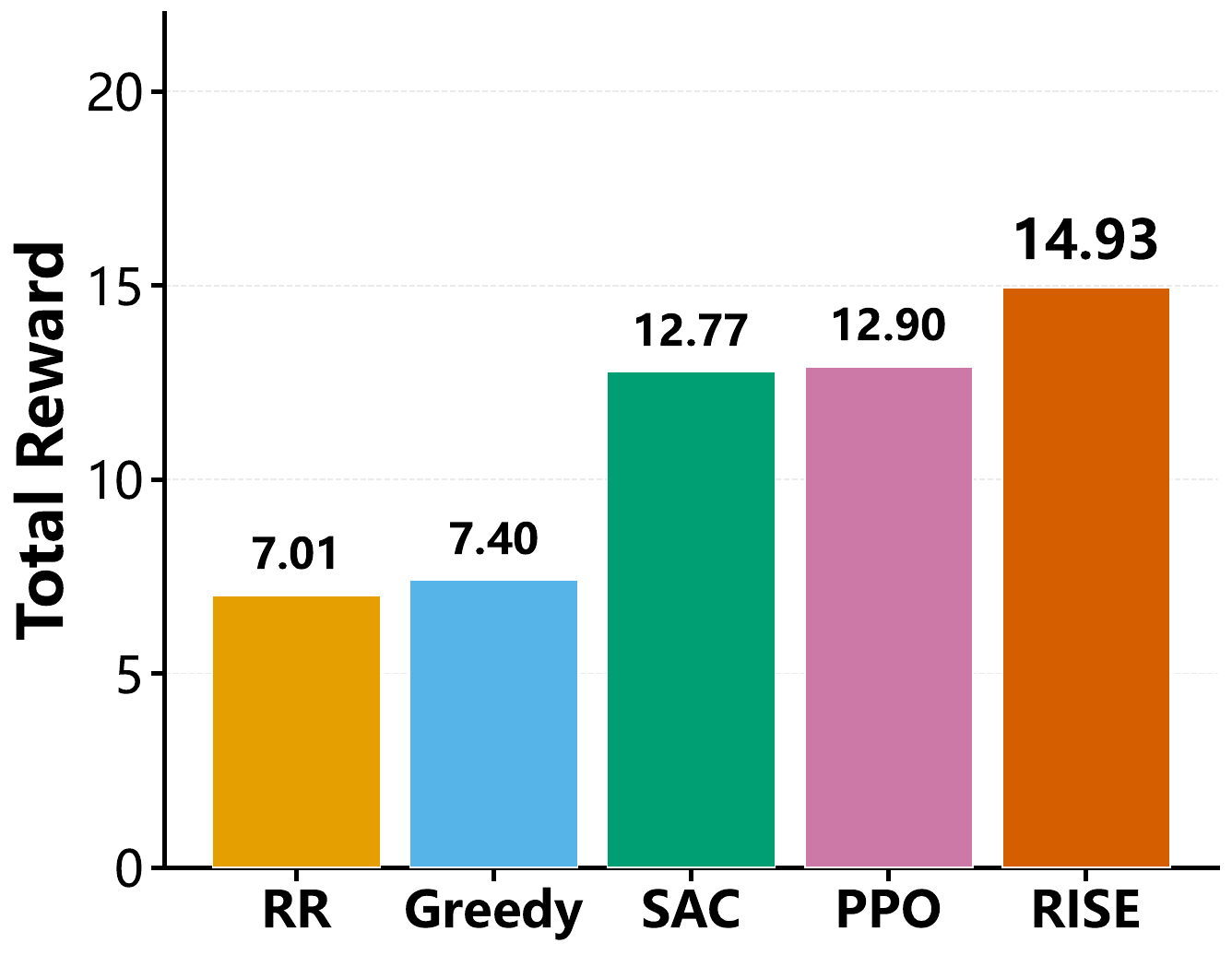}
        \caption{Total Reward}
        \label{fig:total_reward}
    \end{subfigure}%
    \hfill
    \begin{subfigure}{0.22\textwidth}
        \centering
        \includegraphics[width=\textwidth]{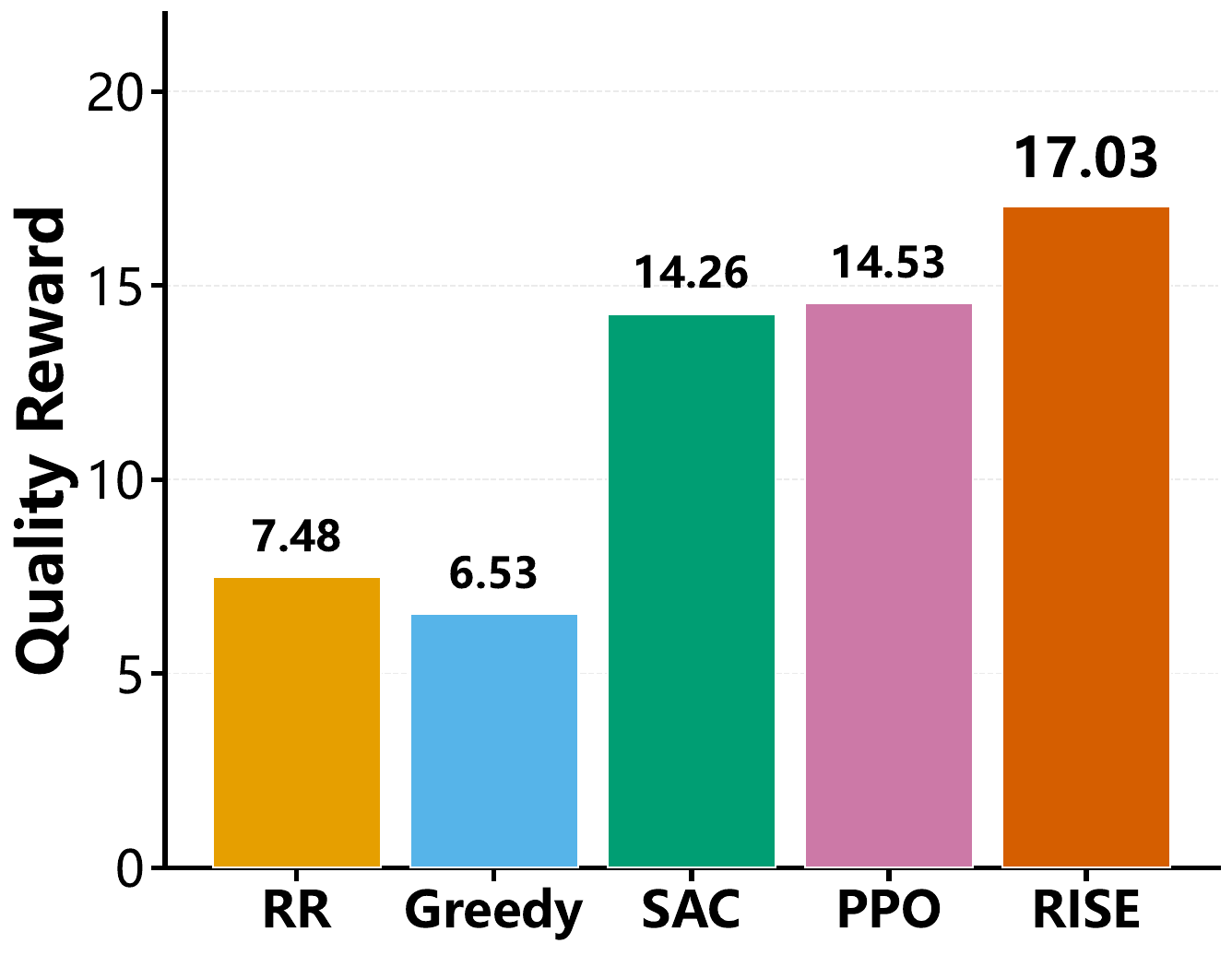}
        \caption{Quality Reward}
        \label{fig:quality_reward}
    \end{subfigure}%
    \hfill
    \begin{subfigure}{0.22\textwidth}
        \centering
        \includegraphics[width=\textwidth]{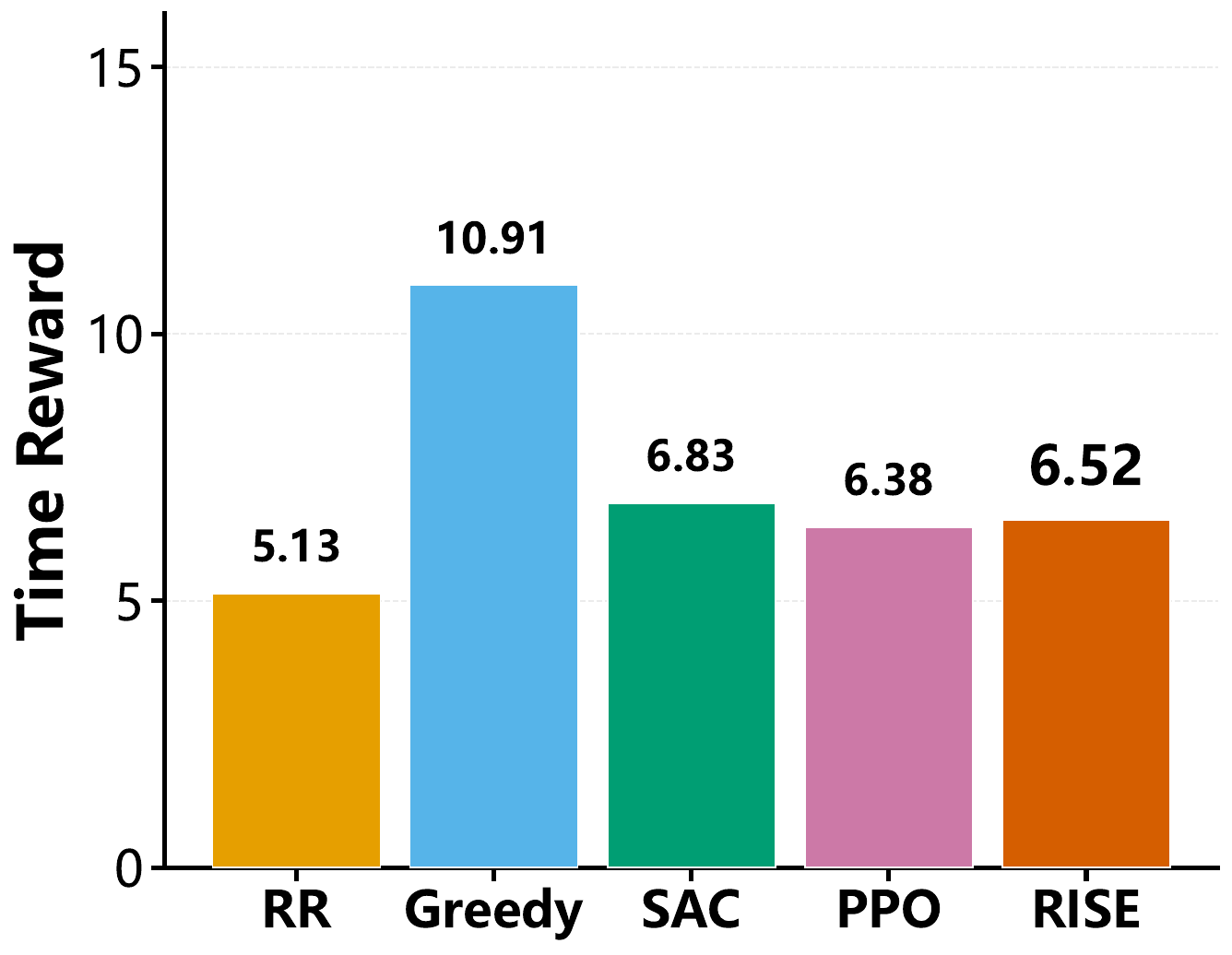}
        \caption{Time Reward}
        \label{fig:time_reward}
    \end{subfigure}%
    \hfill
    \begin{subfigure}{0.25\textwidth}
        \centering
        \includegraphics[width=\textwidth]{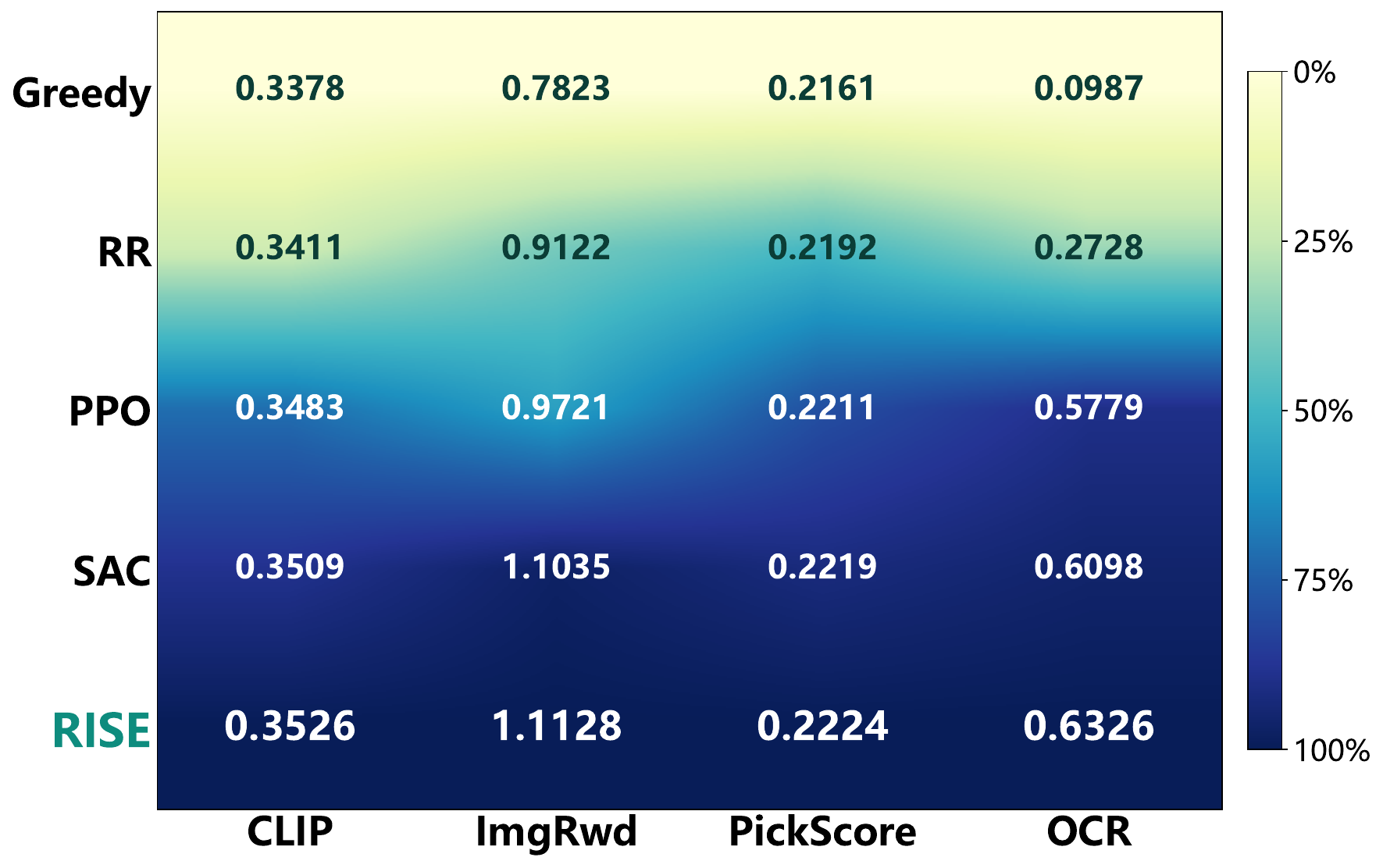}
        \caption{Quality Heatmap}
        \label{heatmap}
    \end{subfigure}
    \caption{Reward and several metrics comparison across five scheduling policies}
    \label{fig:scheduling}
\end{figure*}

\textbf{Reward Comparison.} Figure~\ref{fig:scheduling} presents the comparison across three reward dimensions. RISE achieves the highest total reward, outperforming PPO, SAC, RR, and Greedy. For quality reward, RISE's advantage is most pronounced: RISE's scheduler learns to route text-heavy prompts to SD3 relay while directing simpler prompts to faster SDXL relay. The heatmap in Figure~\ref{heatmap} provides a fine-grained validation of this quality advantage, showing that RISE consistently secures the top position across all four individual quality dimensions: CLIP Score, ImageReward, PickScore, and a substantial lead in OCR Score. This context-adaptive routing emerges naturally from the dynamic reward function (see Eq. (\ref{eq:reward_raw})).

\begin{table*}[t]
\renewcommand{\arraystretch}{1.2}
\caption{Ablation study on RISE scheduling components}
\label{table_ablation}
\centering
\begin{tabular}{l|ccc|cccc}
\toprule
\multirow{2}{*}{\textbf{Variant}} & \textbf{Total} & \textbf{Quality} & \textbf{Time} & \multirow{2}{*}{\textbf{CLIP}$\uparrow$} & \multirow{2}{*}{\textbf{ImageReward}$\uparrow$} & \multirow{2}{*}{\textbf{PickScore}$\uparrow$} & \multirow{2}{*}{\textbf{OCR}$\uparrow$} \\
 & \textbf{Reward}$\uparrow$ & \textbf{Reward}$\uparrow$ & \textbf{Reward}$\uparrow$ & & & & \\
\midrule
\textbf{RISE} & \textbf{14.93} & \textbf{17.03} & \textbf{6.52} & \textbf{0.3526} & \textbf{1.1128} & \textbf{0.2224} & \textbf{0.6326} \\
w/o Context & 7.51 & 6.81 & 10.32 & 0.3350 & 0.8645 & 0.2052 & 0.1495 \\
w/o Dynamic Reward & 7.21 & 6.68 & 9.34 & 0.3255 & 0.8019 & 0.2147 & 0.1265 \\
w/o Forced Exploration & 11.65 & 13.25 & 5.24 & 0.3420 & 1.0783 & 0.2209 & 0.5249 \\
Fixed Relay Step &  10.89 & 12.14 & 5.89 & 0.3420 & 1.0763 & 0.2206 & 0.5099 \\
\bottomrule
\end{tabular}
\end{table*}

Greedy achieves the highest time reward by aggressively selecting the fastest configurations but sacrifices quality, resulting in extremely low total reward. This demonstrates that purely optimizing for latency without considering task-specific quality requirements leads to suboptimal service outcomes. RR performs worst overall because it ignores context entirely, assigning text-rendering tasks to SDXL relay as often as to SD3 relay.

PPO and SAC achieve competitive quality rewards through learned policies, but RISE scores 15.74\%
higher total reward than the strongest baseline (PPO). Under identical offline training conditions, RISE benefits from (1)~\emph{data efficiency}: its closed-form ridge regression update extracts more signal from the same training data than the gradient-based optimization of PPO and SAC; and (2)~\emph{structured exploration}: the UCB confidence bound provides principled exploration that covers the arm space more systematically than the stochastic policies of PPO and SAC. These advantages are particularly relevant in our setting with a moderate number of discrete arms and a compact context vector, where the linear reward model is expressive enough to capture the quality--latency tradeoff.

% \begin{table}[h]
% \renewcommand{\arraystretch}{1.2}
% \caption{Per-metric quality of scheduling policies}
% \label{tab:quality_compare}
% \centering
% \begin{tabular}{l|cccc}
% \toprule
% \textbf{Method} & \textbf{CLIP}$\uparrow$ & \textbf{ImgRwd}$\uparrow$ & \textbf{PickSc}$\uparrow$ & \textbf{OCR}$\uparrow$ \\
% \midrule
% RR      & 0.3411 & 0.9122 & 0.2192 & 0.2728 \\
% Greedy  & 0.3378& 0.7823 & 0.2161 & 0.0987 \\
% SAC     & 0.3509 & 1.1035 & 0.2219 & 0.6098 \\
% PPO     & 0.3483 & 0.9721 & 0.2211 & 0.5779 \\
% \rowcolor{gray!20} \textbf{RISE}    & \textbf{0.3526} & \textbf{1.1128} & \textbf{0.2224} & \textbf{0.6326} \\
% \bottomrule
% \end{tabular}
% \end{table}

\subsection{Ablation Study (RQ3)}

To verify the contribution of each component in the RISE scheduler, we conduct ablation experiments by removing one component at a time and re-running the scheduling evaluation under identical conditions. The results in Table~\ref{table_ablation} reveal clear contributions from each component.

\textbf{(1) Context-aware features} are critical: removing them causes Total Reward to drop from 14.93 to 7.51 ($-49.7\%$), and OCR Score collapses from 0.6326 to 0.1495 ($-76.4\%$), confirming that without prompt-level context the scheduler cannot distinguish text-rendering tasks and fails to route them to appropriate relay configurations.

\textbf{(2) Dynamic reward shaping} is equally important: the \emph{w/o Dynamic Reward} variant suffers the lowest Quality Reward and the worst OCR Score, indicating that fixed reward weights cannot adapt to task-specific quality requirements.

\textbf{(3) Forced exploration} provides a moderate but consistent benefit: without it, Total Reward decreases from 14.93 to 11.65 ($22.0\%$), and the OCR Score drops from 0.6326 to 0.5249 ($17.0\%$), further suggesting that the cold-start exploration phase is beneficial for the bandit to discover high-quality relay arms early.

\textbf{(4) Adaptive relay step selection} further improves the overall efficiency of task scheduling. Compared to the full system, using a fixed relay step reduces the Total Reward by $27.1\%$, as a static split point fails to accommodate the diverse complexity of incoming tasks.

Overall, the context-aware features and dynamic reward shaping contribute the most, while forced exploration and adaptive relay steps provide complementary gains.

\section{Conclusion}
\label{sec:conclusion}

This paper has proposed RISE, a method for collaborative diffusion models across edge servers and user devices. The large edge model runs the early denoising steps that shape semantic structure, and the small device model takes over for detail refinement, requiring no retraining. The relay inference can cut generation time by up to $2.1\times$ while preserving or even improving output quality. A LinUCB-based scheduler further selects the best relay configuration per request by encoding prompt complexity, user preference, network quality and real-time node loads into a context vector, and it scores 15.74\% higher total reward than the strongest baseline across two benchmarks. Currently, RISE requires the large and small models to share the same latent space, limiting relay to within a single model family. A promising direction is to introduce a lightweight adapter that aligns latent representations across families, which enables cross-family relay.

\bibliographystyle{IEEEtran}
\bibliography{refs}

\end{document}